\documentclass[10pt]{article}

\usepackage{amsmath}
\usepackage{amssymb}

\usepackage{graphicx}
\usepackage{cite}

\usepackage{color}

\usepackage[labelfont=bf,labelsep=period,justification=raggedright]{caption}

\makeatletter
\renewcommand{\@biblabel}[1]{\quad#1.}
\makeatother

\date{}

\usepackage{graphicx}
\usepackage{dcolumn}
\usepackage{bm}
\usepackage[english]{babel}
\usepackage[latin1]{inputenc}
\usepackage{graphicx}
\usepackage{epstopdf}
\usepackage{amsfonts}
\usepackage{comment}
\usepackage{color}
\usepackage{epsfig}
\usepackage{mathrsfs}
\usepackage{amsmath,amssymb}
\usepackage{subfigure}
\usepackage{framed}
\usepackage[svgnames]{xcolor}

\renewcommand{\[}{\left[}
\renewcommand{\]}{\right]}
\newcommand{\erf}{\text{erf}}

\newcommand{\beq}{\begin{equation}}
\newcommand{\eeq}{\end{equation}}
\newcommand{\ba}[1]{\begin{array}{#1}}
\newcommand{\ea}{\end{array}}
\newcommand{\bea}{\begin{eqnarray}}
\newcommand{\eea}{\end{eqnarray}}

\newcommand{\ben}{\begin{enumerate}}
\newcommand{\een}{\end{enumerate}}
\newcommand{\bit}{\begin{itemize}}
\newcommand{\eit}{\end{itemize}}
\newcommand{\bde}{\begin{description}}
\newcommand{\ede}{\end{description}}

\newcommand{\ds}{\displaystyle}

\newcommand{\sss}{\scriptscriptstyle}
\newcommand{\sz}{\scriptsize}

\newcommand{\wout}[2]{{#1}_{#2}^{\mbox{\sz out}}}
\newcommand{\win}[2]{{#1}_{#2}^{\mbox{\sz in}}}
\newcommand{\subi}[2]{{#1}_{{\sss {#2}}}}
\newcommand{\supe}[2]{{#1}^{{\sss {#2}}}}

\newcommand{\avg}[1]{\langle {#1} \rangle}

\definecolor{MyBlue}{rgb}{0,0,0}

\begin{document}

\begin{flushleft}
{\Large
\textbf{Emergence of assortative mixing between clusters of cultured neurons}
}
\\
Sara Teller$^{1}$,
Clara Granell$^{2}$,
Manlio De Domenico$^{2}$,
Jordi Soriano$^{1,\ast}$,
Sergio G\'omez$^{2}$,
Alex Arenas$^{2}$
\\
\bf{$^1$} Departament d'Estructura i Constituents de la Mat\`eria, Universitat de Barcelona, Barcelona, Spain
\\
\bf{$^2$} Departament d'Enginyeria Inform\`{a}tica i Matem\`{a}tiques, Universitat Rovira i Virgili, Tarragona, Spain
\\
$^{\ast}$Corresponding author. E-mail: jordi.soriano@ub.edu
\end{flushleft}

\section*{Abstract}
The analysis of the activity of neuronal cultures is considered to be a good proxy of the functional
connectivity of {\em in vivo} neuronal tissues. Thus, the functional complex network inferred from activity patterns is a promising  way to unravel the interplay between structure and functionality of neuronal systems. Here, we monitor the spontaneous self-sustained dynamics in neuronal cultures formed by interconnected aggregates of neurons ({\em clusters}). Dynamics is characterized by the fast activation of groups of clusters in sequences termed bursts. The analysis of the time delays between clusters' activations within the bursts allows the reconstruction of the {\em directed} functional connectivity of the network. We propose a method to statistically infer this connectivity and analyze the resulting properties of the associated complex networks. Surprisingly enough, in contrast to what has been reported for many biological networks, the clustered neuronal cultures present assortative mixing connectivity values, as well as a rich--club core, meaning that there is a preference for clusters to link to other clusters that share similar functional connectivity, which shapes a `connectivity backbone' in the network. These results point out that the grouping of neurons and the assortative connectivity between clusters are intrinsic survival mechanisms of the culture.


\section*{Author Summary}

The architecture of neuronal cultures is the result of an intricate self-organization process that balances structural and dynamical demands. We observe that when the motility of neurons is allowed, these neurons organize into compact clusters. These neuronal assemblies have an intrinsic synchronous activity that makes the whole cluster firing at unison. Clusters connect to their neighbors to form a network with rich spontaneous dynamics. This dynamics ultimately shapes a directed functional network whose properties are investigated using network descriptors. We find that the networks are formed such that preference in connectivity between clusters is based on the similarity between their activity, a property that is called {\em assortative mixing} in networks' language. This particular choice of connectivity correlations must be rooted to basic survival mechanisms for the neurons constituting the culture.

\section*{Introduction}

The theory of complex networks \cite{dorogovtsev2003enb,alon07,Bbarrat08,newman2010networks,estrada} has proven to be a useful framework for the study of the interplay between structure and functionality in social, technological, and biological systems. A complex network is no more than a specific representation of the interactions between the elements of the system in terms of nodes (elements) and links (interactions) in a graph. The analysis of such resulting abstraction of the system, the network, provides clues about regularities that can be connected with certain functionalities, or even be related to organization mechanisms that help to understand the rules behind the system's complexity. Particularly, in biological systems, the characterization of the emergent self-organization of their components is of utmost importance to comprehend the mechanisms of life\cite{Eckmann:2007ft,selforgmigration,Gross1999Origins}.

One of the major challenges in biology and neuroscience is the ultimate understanding of the structure and function of neuronal systems, in particular the human brain, whose representation in terms of complex networks is especially appealing \cite{Bullmore2009186, Sporns2011Human}. In this case, structural connectivity corresponds to the anatomical description of brain circuits whereas the functional connectivity is related to the statistical dependence between neuronal activity.

Network theory and its mathematical framework have provided, through the analysis of the distribution of links, statistical measures that highlight key topological features of the network under study. These measures have facilitated the comprehension of processes as complex as brain development \cite{Sporns2004Organization}, learning \cite{Bassett2011} and dysfunction \cite{Seeley2009,Zhou2012}. Particularly, these measures have unfolded new relationships between brain dynamics and functionality. For instance, synchronization between neuronal assemblies in the developing hippocampus has been ascribed to the existence of super-connected nodes in a scale-free topology \cite{Bonifazi2009}; efficient information transfer has been associated to circuits with small-world features \cite{Latora2001}, such as in the the nematode worm \emph{C. elegans} \cite{watts98} or the brain cortex \cite{Sporns2004SM,Harriger2012}; and the coexistence of both segregated and integrated activity in the brain has been hypothesized to arise from a modular circuit architecture \cite{Hagmann2008,Meunier2010,Gorka2010}.

A network measure that has recently caught substantial attention is the \emph{assortativity coefficient}, which quantifies the preference of a node to attach to another one with similar (\emph{assortative mixing}) or dissimilar (\emph{disassortative mixing}) number of connections \cite{newman2002assortative,newman2003mixing}. Assortative networks have been observed in both structural \cite{Hagmann2008} and functional \cite{Eguiluz2005Scale} human brain networks. It has been  proposed that assortative networks exhibit a modular organization \cite{modul_madrid}, display an efficient dynamics that is stable to noise \cite{Franciscis2011}, and manifest resilience to node deletion (either random or targeted) \cite{Rubinov20101059,newman2002assortative}. Resilience is ascribed to the preferred interconnectivity of high--degree nodes, which shape a `connectivity backbone' \cite{Achard2006} that preserves network integrity. The existence of such a tightly interconnected community is generally known as the `rich--club' phenomenon \cite{Colizza2006RC,Gorka2010,Harriger2012}.
On the other hand, disassortative networks, such as the ones identified in the yeast's protein--protein interaction and the neuronal network of \emph{C. elegans}  \cite{newman2002assortative}, are more vulnerable to targeted attacks. However, in these disassortative networks, the tendency of high degree nodes to connect with low degree ones results in a star--like topology that favors information processing across the network.

The assortativty coefficient is usually calculated through the Pearson correlation coefficient between the unweighted degrees of each link in the network \cite{newman2002assortative}. To account for effects associated to large networks, the Spearman assortativity measure was introduced \cite{spearman2013} and, later, weighted assortativity measures were proposed to include the weight in degree--degree dependencies \cite{Leung2007591}.

To better understand the importance of these network measures in describing neuronal networks, \emph{in vitro} preparations in the form of neuronal cultures have been introduced given their accessibility and easy manipulation \cite{Eckmann:2007ft,Wheeler2010Designing}. Two major types of cultured neuronal networks are of particular interest. In a first type, neurons are plated on a substrate that contains a layer of adhesive proteins. Neurons firmly adhere to the substrate, leading to cultures with a homogeneous distribution of neurons  \cite{Wagenaar:2006fo,Cohen200821,Orlandi2013np,Tibau2013}. In a second type, neurons are plated without any facilitation for adhesion. Neurons then spontaneously group into small, compact assemblies termed \emph{clusters} that connect to one another \cite{teller:244,Gabay2005Enginereed,Segev2003Electrically,SheinIdelson2010Innate}.

The formation of a clustered architecture from an initially isotropic configuration is an intriguing self-organization process\cite{Segev2003Electrically,SheinIdelson2010Innate}. This feature has made clustered networks attractive platforms to study the development of neuronal circuits as well as the interplay between structural and functional connectivity at intermediate, {\em mesoscopic} scales \cite{Gabay2005Enginereed,Soussou2007, Sorkin2006Compact,Macis2007Microdrop ,Shein2009Engineered}. Moreover, the existence of a two-level network, one within a cluster and another between clusters, has made clustered cultures appealing to study dynamical and topological features of hierarchical \cite{Segev2003Electrically, Shein2009Engineered, SheinIdelson2010Innate, SheinIdelson2011Modular} as well as modular networks \cite{Tsai2008Robustness, Yvon2005Patterns, Berdondini2006530, SheinIdelson2011Modular}.

In this work we use spontaneous activity measurements in clustered neuronal cultures to render the corresponding directed functional networks and study their topological properties. We introduce a novel theoretical framework that uses the propagation of activity between clusters as a measure of ``causality'', although strictly speaking we should refer to as a sequence of delayed activations, giving rise to functional connections that are both directed and weighted. Based on this weighted nature of the network, we propose a new measure of assortativity that explicitly incorporates the weight of the links. We observed that all the studied functional networks derived from clustered cultures show a strong, positive assortative mixing that is maintained along different stages of development. On the contrary, homogeneous cultures tend to be weakly assortative, or neutral. Finally, in combination with experiments that measure the robustness of network activity to circuitry deterioration, we show that the strongly assortative, clustered networks are more resistant to damage compared to the weakly assortative, homogeneous ones. Our work provides a prominent example of the existence of assortativity in biological networks, and illustrates the utility of clustered neuronal cultures to investigate topological traits and the emergence of complex phenomena, such as self-organization and resilience, in living neuronal networks.

\section*{Results}

\subsection*{Experiments}

\begin{figure}[!b]
\begin{center}
\includegraphics[width=16cm]{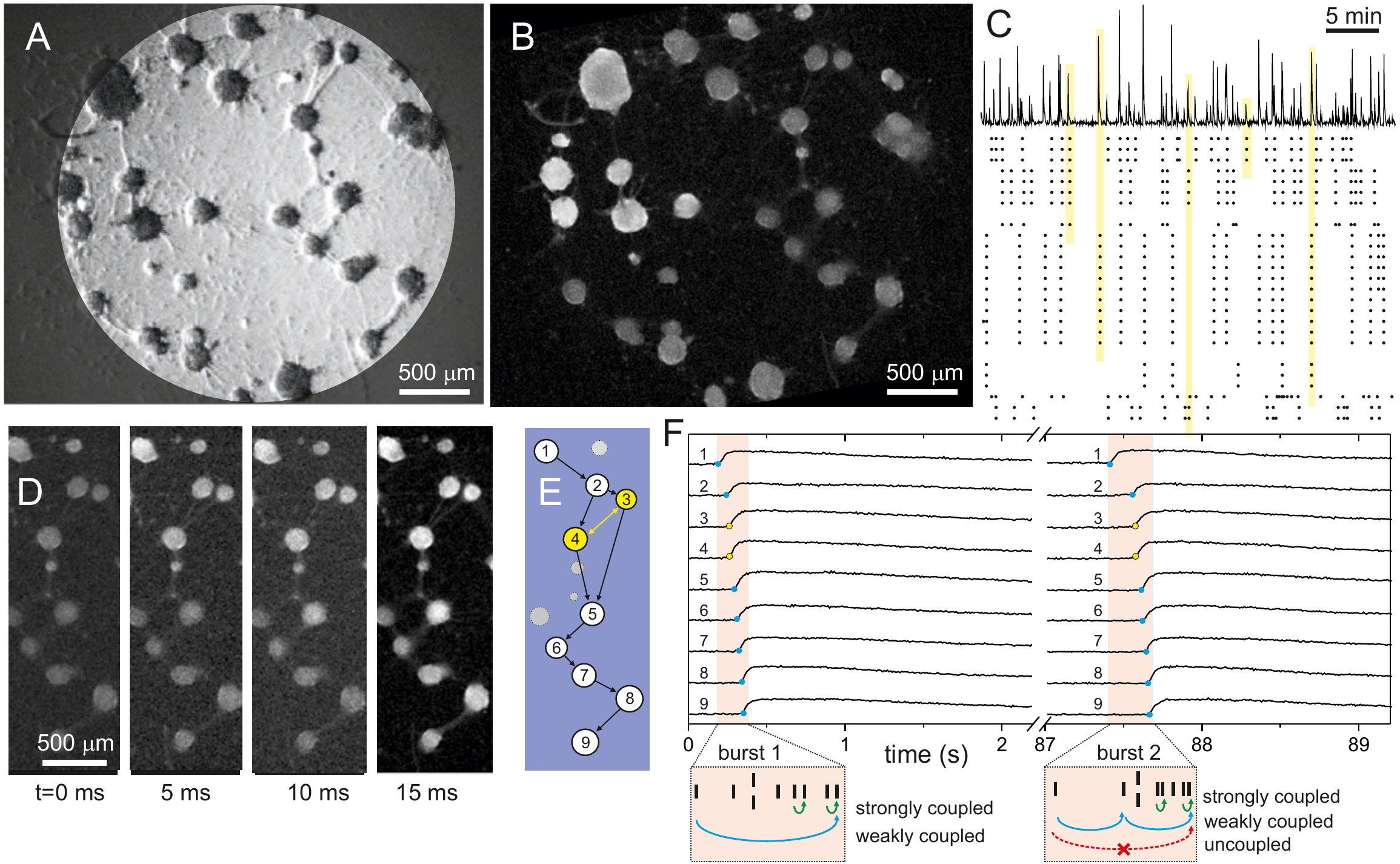}
\end{center}
\caption{\textbf{Experiments in clustered neuronal networks.}
\textbf{A} Bright field image of a network at \emph{day in vitro} $14$. Dark circular objects are aggregates of neurons (clusters), and filaments are visible physical connections between them.
\textbf{B} Corresponding fluorescence image, integrated over 50 frames ($\approx 0.5$~s). Bright clusters at the top-left corner are active ones.
\textbf{C} Spontaneous activity in the network (see Movie S1 for the actual recording). The top plot shows the average fluorescence signal of the clustered network shown in \textbf{B}, and along 40~min of recording. The sharp peaks in fluorescence correspond to the fast sequential ignition of a group of clusters (burst). The bottom raster plot shows the clusters that ignite along the recording. The yellow bars relate a fluorescence peak with the ignition of a group of clusters, and highlights the tendency for the clusters to activate in specific groups.
\textbf{D} Example of a particular ignition sequence in a region of the network containing $13$ clusters. From left to right, the progress of cluster's activation is revealed by the increase in fluorescence signal of the downstream connected clusters.
\textbf{E} Order of activation (black arrows) according to the analysis of the fluorescence signal. The clusters marked in yellow are those that fire simultaneously within experimental resolution. The ones in grey are clusters that do not participate in the firing sequence, and either fire independently or remain silent.
\textbf{F} Detail of the fluorescence traces for the $9$ participating clusters along two consecutive bursts, illustrating the accuracy in resolving the time delay in the activation of the clusters. The two bursts contain the same clusters, but the activation sequences are slightly different. Blue dots mark the ignition time, and yellow dots signal the clusters that fired simultaneously. The bottom orange boxes depict the final activation sequences of each burst. In the construction of the directed functional network, the influence of a cluster on another is conditioned by the time span between their activations. Close activations result in strong couplings (green arrows); far activations in weak ones (blue). Any two clusters whose activations are above $200$~ms are considered functionally uncoupled (red).}
\label{fig:experiments}
\end{figure}

We used rat cortical neurons in all the experiments. As described in Methods, neurons were dissociated and seeded homogeneously on a glass substrate. Cultures were limited to circular areas 3~mm in diameter for better control and full monitoring of network behavior. The lack of adhesive proteins in the substrate rapidly favored cell--to--cell attachment and aggregation, giving rise to \emph{clustered} cultures that evolved quickly (Figure~\ref{fig:experiments}A). By \emph{day in vitro} (DIV) 2, cultures contained dozens of small aggregates, which coalesced and grew in size as the culture matured. Connections between clusters as well as initial traces of spontaneous activity were observed as early as DIV $4$. Cultures comprised of $20-40$ interconnected clusters by DIV $5$, and were sufficiently stable and rich in activity for measurements. Although the strength of the connections in the network and its dynamics evolved further, we observed that the size and position of the clusters remained stable. We therefore measured dynamics already at DIV $5$, and studied cultures up to DIV $16$.

The example shown in Figure~\ref{fig:experiments}A corresponds to a culture at DIV $14$. Clusters appear as circular objects with an average diameter of $200~\mu$m and a typical separation of $250~\mu$m. Connections between clusters are visible as straight filaments that contain several axons.

We monitored spontaneous activity in the clustered network through fluorescence calcium imaging (Figure~\ref{fig:experiments}B). Fluorescence images of the clustered network were acquired at a rate of 83--100 frames per second, and with an image size and resolution that allowed the monitoring of all the clusters in the network with sufficient image quality (see Movie S1). Activity was recorded for typically 1 hour, which provided sufficient statistics in firing events while minimizing culture degradation due to photo-damage.

The analysis of the images at the end of the measurement provided the variations in fluorescence intensity for each cluster and the corresponding onset times of firing (see Methods). As shown in the top panel of Figure~\ref{fig:experiments}C, the average fluorescence signal of the network is characterized by peaks of intense cluster activity combined with silent intervals. The accompanying raster plot reveals that this activity actually corresponds to the collective ignition of a small group of clusters, which fire sequentially in a short time window on the order of few hundred milliseconds. We denote by \emph{bursts} these fast sequences of clusters' activations. Young cultures (DIV $5-7$) exhibited an activity of about $5$ bursts/min, while maturer cultures (DIV $8-14$) displayed about $2$ bursts/min (see also Table~\ref{tab:results}).

We observed that the time spanned between two consecutively firing clusters typically ranged between $10$ and $100$ ms (see Methods), as also observed by others \cite{Tsai2008Robustness,Yvon2005Patterns}. These times are fairly large compared to the eventual scale of signal integration--propagation between single neurons ($\approx 5$ms), and is related to the large time scales associated to integration of the intra--clusters information. No consecutive activations were observed above $200$ ms, signaling the termination of a burst. We therefore use this value of $200$~ms as a cut--off to separate a given burst form the preceding one. Then, two clusters that fired above 200 ms cannot be influenced by one another and therefore are not causally connected.

Bursts occurred every 30 s on average for the experiment shown in Figure~\ref{fig:experiments}C and, as illustrated by the yellow bands in this figure, each burst typically encompassed a subset of clusters rather than the entire network. In general, however, the number of participating clusters within a burst depended on the details of the culture. Although in a typical experiment the collective firing comprised between $2$ and $10$ clusters (see Movie S1), in some experiments the entire cluster population lighted up in a single bursting episode.

The analysis of the onset times of firing provides the cluster's activation sequence within each burst. As an example, Figure~\ref{fig:experiments}D depicts a highly active region of the network shown in Figure~\ref{fig:experiments}B. This region contains 13 clusters, and 9 of them form a subset that regularly fires together. The series of frames show the progress in clusters' activation, revealed by the changes in fluorescence. Activity starts at the top-left cluster and progresses downwards. The time-line of sequence activation after image analysis is shown in Figure~\ref{fig:experiments}E, and the actual fluorescence traces are shown in Figure~\ref{fig:experiments}F. With our 10 ms temporal resolution we could resolve well the propagation of activity from a cluster to its neighboring ones (black arrows in Figure~\ref{fig:experiments}E). However, and for about 5\% of the cases, the time delay between clusters' activation was either too short for detection or activation occurred simultaneously. The clusters associated to these `simultaneous' events are marked in yellow in Figure~\ref{fig:experiments}E, and their inter--relation was treated as a bi-directional link (yellow arrow), since no causality can be inferred.

A typical recording provided on the order of $100$ bursting episodes. Some of them included the same group of clusters, although the precise sequence of activation could vary. An illustrative example is shown in Figure~\ref{fig:experiments}F, which depicts the fluorescence traces of $9$ clusters along two consecutive bursts. The first sequence corresponds to the sketch of Figure~\ref{fig:experiments}E. The orange box at the bottom of the plot indicates the relative activation time of each cluster within the window, with two clusters treated as simultaneous. To introduce the construction of the directed functional network that is described later, we note that, intuitively, the firing of cluster \#9 is most likely caused by \#8 and therefore both clusters are (functionally) strongly coupled. At the other extreme, cluster \#1 most likely did not trigger \#9, and therefore their mutual coupling is very weak. For the second burst, we note that the activation sequence is very similar, but the relative delay times differ, therefore modifying the cluster's coupling strengths. Indeed, cluster \#1 and \#9 are now functionally disconnected given their long temporal separation.

We carried out measurements in $15$ different clustered networks, and labeled them with capital letters as networks $\text{`A'}-\text{`O'}$. In order to compare their properties with the ones from cultures with a distinct structure, we applied the same measuring protocols and data analysis to $6$ cultures characterized with a homogeneous distribution of neurons (see Methods and Figure S1), and labeled them as networks $\text{`P'}-\text{`U'}$.

\begin{figure}[!t]
\begin{center}
\includegraphics[width=12cm]{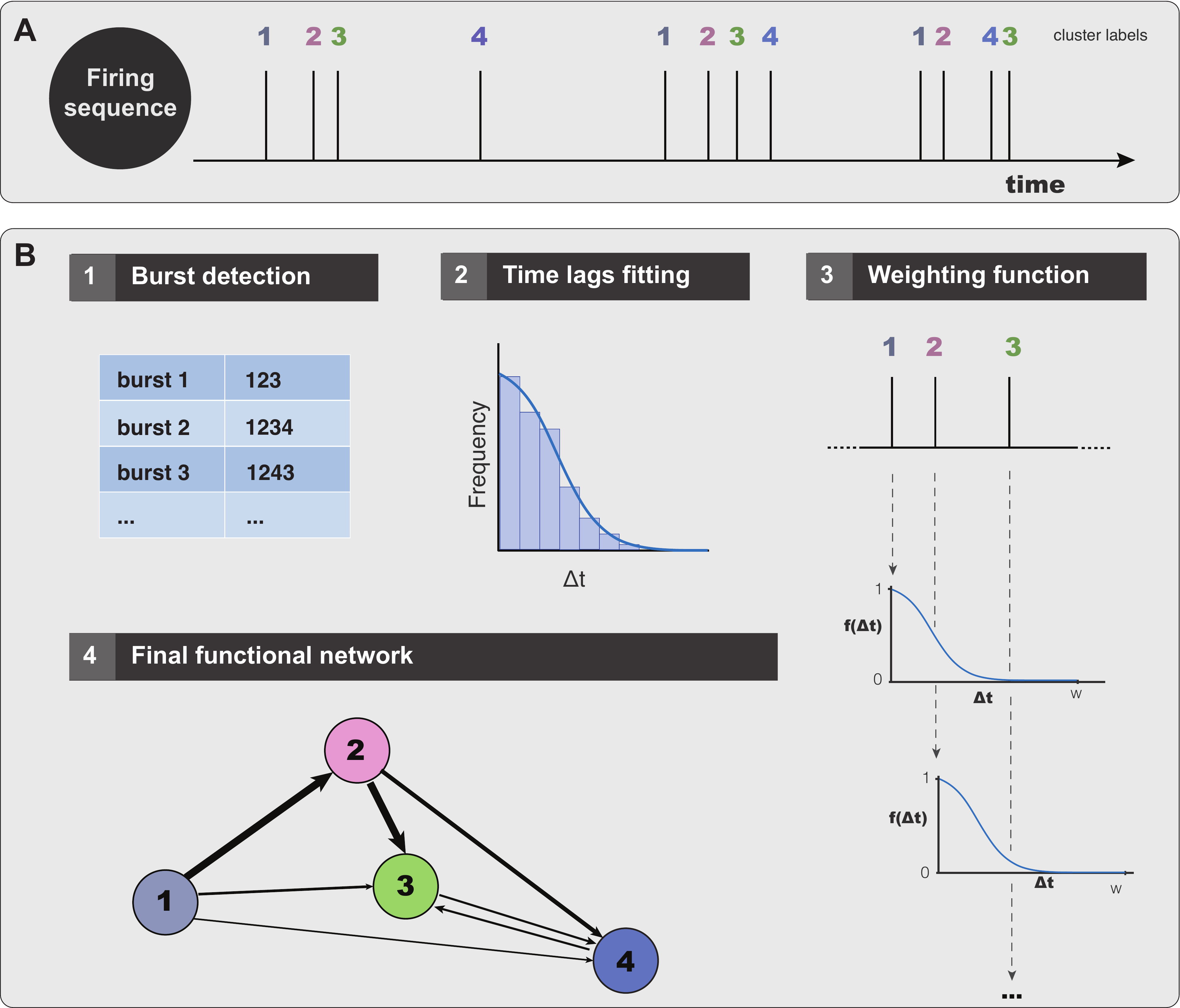}
\end{center}
\caption{\textbf{Sketch of the construction of the directed functional network.}
\textbf{A} Schematic representation of the experimental data, with 12 firings of four different clusters (or neurons in the case of homogeneous cultures).
\textbf{B} Stages of the method to construct the directed functional network.
(1) The first step consists in detecting the different bursts in the whole sequence. Firings that are separated by more than 200 ms for clustered cultures (10 ms for homogeneous ones) are not considered part of the same burst. (2) Calculation of the time lags $\Delta t$ between consecutive firings inside the bursts, for the whole sequence. The frequency distribution of time lags is next fitted to a Gaussian distribution $\text{f}(\Delta t) \sim \text{exp}(-(\Delta t)^2/c)$, finally providing the variance $c$ that will be specific for each culture. (3) Weighting procedure example for the first burst. Cluster \#1 can activate \#2  and \#3. The weight of the links 1$\rightarrow$2 and 1$\rightarrow$3 depends on the time differences between clusters' activations, and is given by the function $\text{f}(\Delta t)$ determined by the previous fitting. Hence, weight $w_{1\rightarrow2} \simeq 0.6$, and  $w_{1\rightarrow3} \simeq 0$. Cluster \#3 can be activated as well by cluster \#2, with $w_{2\rightarrow3} \simeq 0.2$. (4) Schematic representation of the resulting directed functional connectivity network. The width of the connections is proportional to the weight of the links for clarity.}
\label{fig:sketch}
\end{figure}

\subsection*{Construction of the directed functional networks}

The above sequences of clusters' activations, extended to all the clusters and bursting episodes of the monitored culture, convey information on the degree of causal influence between any pair of clusters in the network. For instance, cluster \#5 in Figures~\ref{fig:experiments}E-F can fire because of the first order influence of clusters \#3 and \#4, but also because of the second and third order influences of clusters \#2 and \#1, respectively. Hence, a realistic functional network construction should take into account these possible influences from the upstream connected clusters to build a network whose links are not only directed, but also weighted by the time delays in activation. This weighted treatment of the interaction between clusters is the major novelty of our work and the backbone of our model.

More formally, the interaction between any two clusters follows the principle of causality, i.e. the firing of cluster $j$ immediately after cluster $i$ eventually implies that cluster $i$ has induced the activity of $j$ at that particular time. The likelihood of this relation between clusters is weighted according to its frequency along the full observational time, allowing to a statistical validation. Indeed, cluster $i$ could induce the activity of various clusters, if all of them activate in a physically plausible short time window after cluster $i$. Such a construction is illustrated in Figure~\ref{fig:sketch}.

To construct the directed functional networks for each studied culture we proceed as follows. First of all, we divide the entire firing sequence into the bursts of clusters' activity (Figure~\ref{fig:sketch}A) using the cut--off of $200$ ms introduced in the previous section. Once the bursts have been detected, we compute the frequency distribution of time lags between pairs of consecutive firings (Figure~\ref{fig:sketch}B). This frequency distribution informs about the characteristic times expected between two consecutive firings within the same burst, and hence it is a good proxy of the causal influence of a cluster on another. We will use this information to weight the causal influence of firing propagation. The frequency distribution $\text{f}(\Delta t)$ presents a good fit to a universal Gaussian decay ($y\sim e^{-x^{2}/c}$) in all the analyzed cultures, although the variance $c$ is specific for each culture. We indeed observed (Figure S2) that $c$ decreases with the culture age {\em in vitro} (correlation coefficient $R=-0.81$, significance $p=0.008$), and increases with the number of clusters present in the network ($R=0.62$, $p=0.03$).

The last step in the construction of the directed functional networks consists in linking the interactions within each burst, and weighting them according to the previous frequency distribution (Figure~\ref{fig:sketch}B). The rationale behind this process is as follows: we hypothesize that every cluster influences other clusters (posterior in time) within a burst and, the larger the time after a cluster has fired the lower the influence we expect in the activation of another cluster (simply because the signal fades out). Then, the weighting of the interaction by the expected frequency observed in the distribution conveys the functional influence between clusters. The weights are reinforced every time the same pair of clusters' sequence is observed. After processing the full sequence we obtain a peer-to-peer activation map that is our proxy of the functional network.

We proceeded identically to construct the directed functional networks for homogeneous cultures (see Methods), with the only difference that the cut-off time corresponds to $10$ ms. We tested for both clustered and homogeneous cultures that the obtained functional networks were stable upon variations of the cut-off times (see Figure S3 and Discussion).

\subsection*{Analysis of the functional networks}

\begin{figure}[!t]
\begin{center}
\includegraphics[width=16cm]{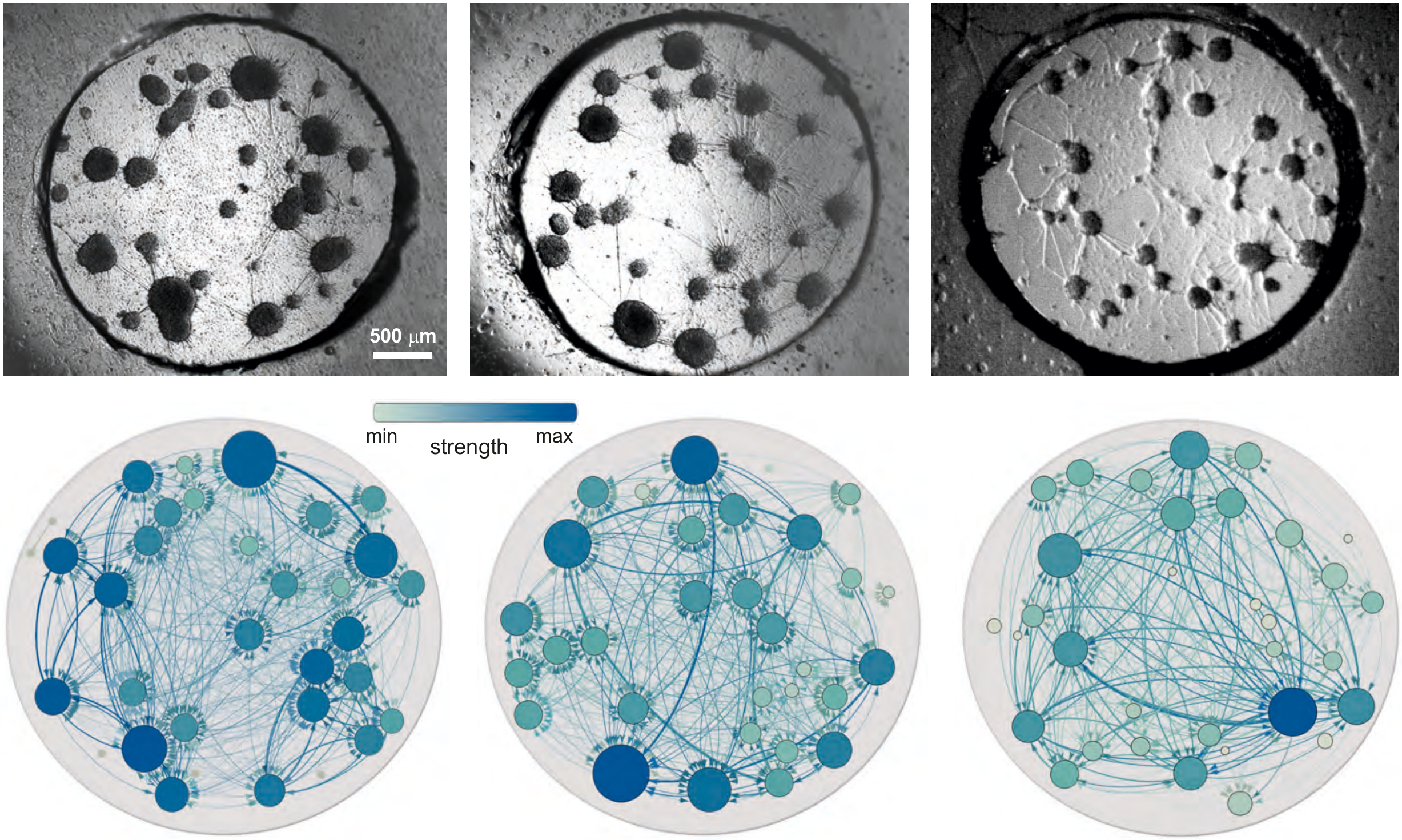}
\end{center}
  \caption{\textbf{Neuronal cultures and functional networks.} Top: Bright field images of $3$ representative neuronal cultures at different \emph{days in vitro}. Bottom: Corresponding functional networks obtained from the directed and weighed construction described in Figure~\ref{fig:sketch}. From left to right, the pictures correspond to the cultures labeled D, H and O in Table~\ref{tab:results}.  Only active clusters are used in the construction of the functional network. The size of the nodes is similar to the ones observed in the cultures, and facilitates the comparison of the functional network with the real culture. In the functional networks, the edges are both color and thickness coded according to their weight, while the nodes are only color coded according to their strength. The darker the color, the higher the value.}
  \label{fig:networks}
\end{figure}

We computed the functional networks of the $15$ (`A' to `O') realizations of clustered cultures, as well as the $6$ (`P' to `U') homogeneous ones, and analyzed some major topological traits. Firstly, for each culture we obtained the number of nodes, the number of edges, the average degree of the networks, and its average strength (see Methods).  The investigated networks and their topological measures are summarized in Table~\ref{tab:results}. Although young cultures display a richer activity, in general all networks present a similar number of nodes and a comparable functional connectivity, which is described by the number of edges, the average degree and the average strength.

\begin{table}[!t]
  \centering
    \begin{center}
  \begin{tabular}{c}
            \includegraphics[width=15.5cm]{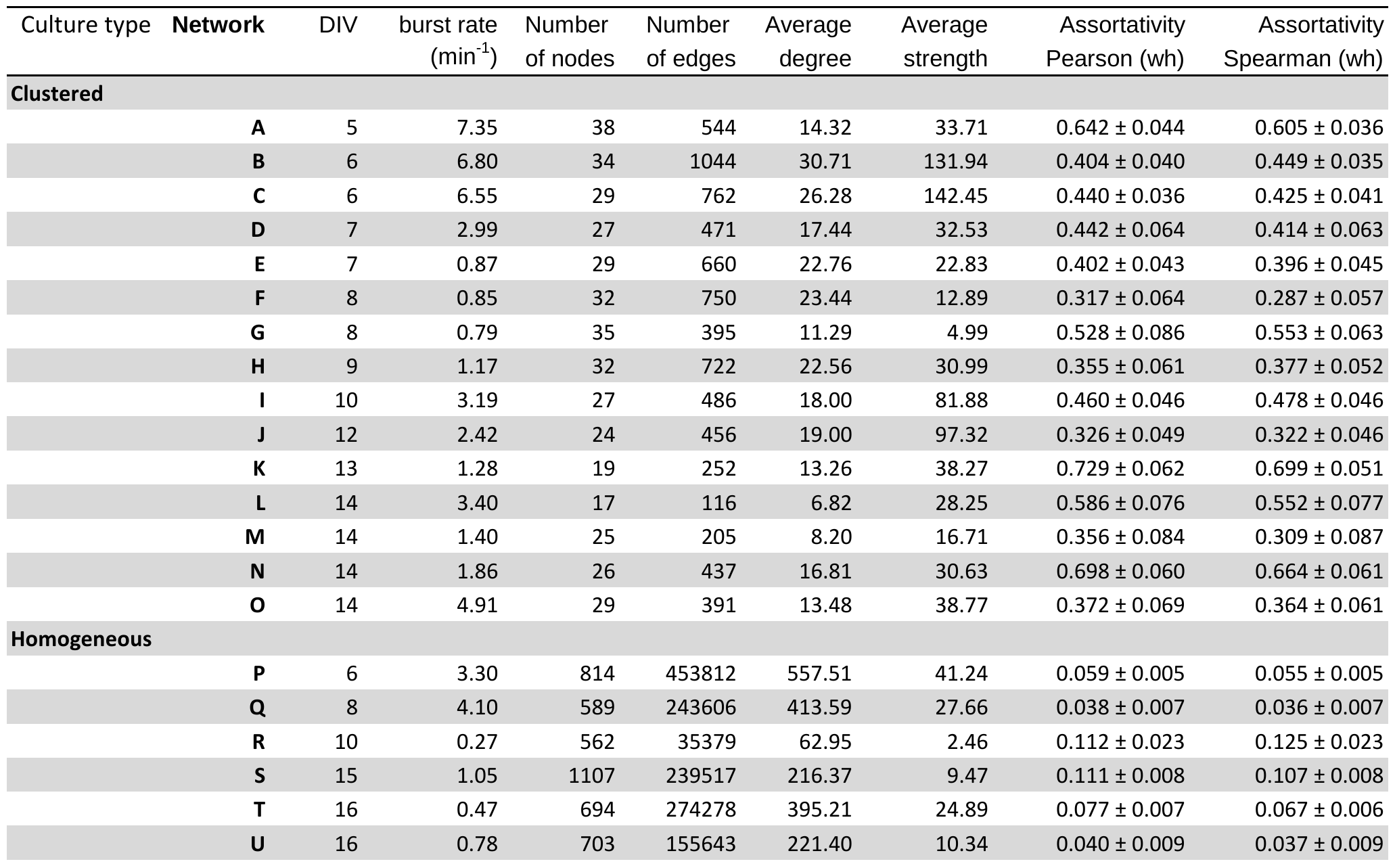}
  \end{tabular}
        \end{center}
  \caption{\textbf{Network measures of clustered and homogeneous cultures}. The table shows the major topological descriptors of the functional networks investigated, $15$ corresponding to a clustered neuronal organization and $6$ to a homogeneous one.  Average degree refers to the mean number of connections per node, and the average strength to the mean weight per node. All the cultures were maintained and studied identically (see Methods). Experiments cover almost $10$ days of development \emph{in vitro} (DIV). All clustered cultures are strongly assortative, while the homogeneous ones tend to be weakly assortative, or neutral.}
    \label{tab:results}
\end{table}

Representative examples of the investigated functional networks for the clustered configuration are shown in Figure~\ref{fig:networks} (see Figure S1 for an example of the homogeneous ones). The position of the nodes and their size are the same as the actual clusters for easier comparison. Edges in the directed network are both color and thickness coded to highlight their importance, with darker colors corresponding to the highest weights. This representation reveals those pairs of clusters that maintain a persistent causality relationship over time. Nodes are also color coded according to their strength, i.e. the total weight of the in-- and out--edges.

The functional networks exhibit some interesting features. First, there are groups of nodes that form tightly connected communities. These communities actually reflect the most frequent bursting sequences. Second, nodes preferentially connect to neighboring ones with some long--range connectivity, and often following paths that are not the major physical connections. This indicates that the structural connectivity of the network cannot be assessed from just an examination of the most perceivable processes. And third, as shown in Figure S4A, we observed that there is no correlation between the width of the physical connections and their weight ($R=0.040$, $p=0.88$), or the size of the nodes and their strength ($R=0.072$, $p=0.70$, Figure S4B), and indicates that the dynamical traits of the network cannot be inferred from its physical configuration, stressing the importance of the functional study.

We also observed that the size of the clusters did not correlate with their average activity ($R=0.14$, $p=0.51$, Figure S4C), i.e. small and big clusters displayed similar firing frequencies, and of $1$ firing/min on average. However, since some clusters are initiators of activity and others just followers, we also computed the relative contribution of a given cluster size to initiate activity in the network. We found no significant correlation between initiation and cluster size ($R=0.38$, $p=0.14$, Figure S4D). These results strengthen the conclusion that one cannot predict the clusters that will initiate activity, or the most persistent sequences, by just a visual inspection of cluster sizes and their distribution over the network.

These analyses are important in the context of the work by Shein--Idelson and coworkers \cite{SheinIdelson2010Innate}, who studied the dynamics of {\it isolated} clusters similar to ours, and observed that their firing rate increased from $0.7$ to $8$ firings/min as the clusters' radii escalated from $30$ to $130$ $\mu$m. This remarkable difference in the dynamics between `isolated' and `networked' clusters reflects the dominant role of the network circuitry in shaping its dynamics.

Finally, to crosscheck that the results found for the functional networks presented here are robust to the inference method, we have also performed a classical mutual information analysis to construct functional networks for the same cultures (see Methods). The results obtained with the mutual information analysis are totally in agreement with the constructed functional networks using time delays.

\subsection*{Assortativity and rich--club properties}

We determined the values of the weighted formulation of assortativity, both for the Pearson $\supe{\rho}{PW}$ and Spearman $\supe{\rho}{SW}$ correlations, with values in $\[-1, 1\]$ (see Methods for the generalization of assortativity to directed weighted networks). Positive values of the weighted assortativity indicate that nodes with similar strength tend to connect to one another, while negative values mean the preferred interconnectivity of nodes with different strength. In Table \ref{tab:results} we can observe that all clustered networks (labeled `A'-`O') exhibit a positive weighted assortativity, in the range $0.32 \leq \supe{\rho}{PW} \leq0.73$ for the Pearson construction and $0.29 \leq \supe{\rho}{SW} \leq0.70$ for the Spearman one. Although the values fluctuate across different cultures, the two assortativity measures provide the same value within statistical error, and reflect that network size corrections provided by the Spearman's treatment have little influence in strongly assortative networks.

To assess the importance of the measured assortativity values, we have also computed the weighted rich--club \cite{Zlatic2009}. The rich--club phenomenon refers to the tendency of nodes with high degree to form tightly interconnected communities, compared to the connections that these nodes would have in a null model that preserves the node's degree but otherwise is totally random. Given the positive assortativity found, we analyzed whether this finding is also reinforced by the existence of rich--club structures.

The weighted formulation for the rich--club takes into account the node's strength instead of the degree, and is particularly useful in situations in which the weights of the links can not be overlooked~\cite{mangels_rich}. The evaluation of the rich--club $\phi^{\text{unc}}(s_T) $ is performed by computing the ratio between the connectivity strength of highly connected nodes and its randomized counterpart, and for gradually higher values of the strength threshold $s_T$. The detailed calculation is described in the Methods section, and the results of the analysis for representative networks is shown in Figure S5. Ratios larger than $1$ indicate that higher strength nodes are more interconnected to each other than what one would expect in a random configuration. On the contrary, a ratio less than $1$ reveals an opposite organizing principle that leads to a lack of interconnectivity among high--degree nodes. After the calculation of the ratios for all the studied clustered networks, we found a positive tendency towards the creation of rich--clubs in all of them (Figure S5), which is in good agreement with the observed values of assortativity.

The above network measures were also analyzed in experiments with a homogeneous distribution of neurons (labeled $\text{`P'}-\text{`U'}$). The results are summarized in Table~\ref{tab:results}. Interestingly, the assortativity values are much lower (by an order of magnitude on average) than the ones for clustered cultures, in the range $0.04 \leq \supe{\rho}{PW} \leq0.11$ for Pearson's and $0.04 \leq \supe{\rho}{SW} \leq 0.12$ for Spearman's. Accordingly, the rich-club phenomenon for the homogeneous cultures vanishes (Figure S5).

\subsection*{Network resilience}

Several studies highlight the importance of assortative features for network resilience to damage. Given the strong assortativity of our clustered cultures, we carried out a new set of experiments to investigate the concurrent presence of resilient traits. As described in Methods, we considered two major `damaging' actions to the network. In a first one, we gradually weakened the excitatory network connectivity by means of the AMPA--glutamate antagonist CNQX, and measured the decay in spontaneous activity as connectivity failed. In a second one, we continuously exposed a culture to strong fluorescence light, therefore inducing photo-damage to the neurons. This action resulted in random neuronal death across the network and hence a progressive failure of its spontaneous dynamics. The rate of activity decay upon radiation damage provided an estimation of the resistance of the network to node deletion. These investigations were carried out at the same time in clustered cultures (strongly assortative) and in homogeneous ones (weakly assortative or neutral). Their comparison provided a first reference to relate assortativity, network topology and resistance to damage.

\begin{figure}[!t]
\begin{center}
\includegraphics[width=16cm]{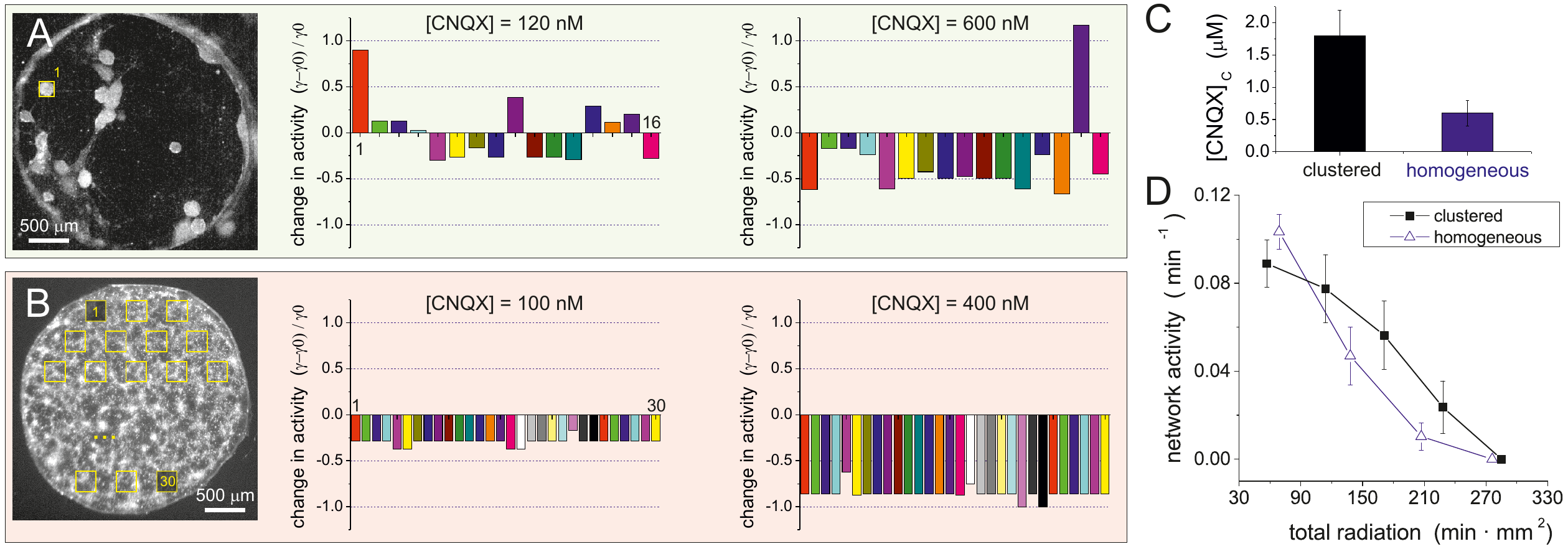}
\end{center}
\caption{\textbf{Network resilience to damage.}
  \textbf{A}--\textbf{B} Examples of the degradation of neuronal activity in clustered and homogeneous cultures due to the gradual weakening of excitatory connectivity. Both culture types were investigated at the same \emph{day in vitro} $14$ and contained a similar density of neurons. The weakening of connections is achieved by gradually increasing the concentration of CNQX, an AMPA-glutamate receptor antagonist in excitatory neurons. Network response upon weakening is quantified through the relative change in activity $(\gamma-\gamma_0)/\gamma_0$ between a given CNQX application and the unperturbed state. Activity variations are indicated separately for each cluster, and shown according to the cluster labeling number.
\textbf{A} Clustered cultures show a mixed response upon weakening, with some clusters increasing activity and others reducing it. Only for relatively high concentrations of CNQX ($\gtrsim 600$ nM) the activity systematically decays up to the full silencing of the network.
\textbf{B} In homogeneous cultures, activity is analyzed in $30$ regions that cover in a regular manner the entire network. Activity decays almost equally in all regions. Relatively small drug concentrations of [CNQX]$\simeq 400$ nM practically suffice to fully stop activity.
\textbf{C} Average critical concentration [CNQX]$_\textbf{C}$ at which spontaneous activity completed ceases, about $1.6\mu$M for clustered networks and $0.5\mu$M for homogeneous ones. Data is averaged over $4$ network realizations of each type of culture.
\textbf{D} Photo--damage experiments. Spontaneous activity is measured in cultures that are continuously exposed to strong fluorescence light, causing gradual neuronal degradation and ultimately the death of the entire network. The total radiation received by the neurons is calculated as the duration of the exposure times the area covered by the neurons in the culture ($1.9$ mm$^2$ and $2.3$ mm$^2$ on average for clustered and homogeneous clusters, respectively). The spontaneous activity in homogeneous cultures decays at a much faster rate than in the clustered counterparts. Data is averaged over $6$ network realization of each type. Error bars show standard deviation.}
\label{fig:resilience}
\end{figure}

Figure~\ref{fig:resilience}A shows the results for the application of CNQX to clustered cultures. We first monitored each cluster individually in the unperturbed case, and measured its average firing activity $\gamma_0$ along $15$ min. We then applied a given drug concentration, measured the firing activity $\gamma$ for another $15$ min, and computed the relative changes in activity respect to the unperturbed case, as $\Gamma \equiv (\gamma-\gamma_0)/\gamma_0$. The protocol was repeated until activity ceased. Two illustrative examples of the action of CNQX on network activity are provided in Figure~\ref{fig:resilience}. In a clustered cultured and for weak CNQX applications ($\simeq 100$nM) the activity in some clusters increases, while in some other decreases, and on average the network firing rate remains stable ($\Gamma \simeq 0$). As CNQX is increased to $600$ nM, we observe that most of the clusters have reduced their activity, although there are still some that maintain a high activity or even increase it. This different behavior from cluster to cluster suggests that clustered networks are highly flexible, and that they may have mechanisms to preserve activity even with strong weakening of the connectivity. Conversely, homogeneous cultures (Figure~\ref{fig:resilience}B) lose activity in a more regular and faster way. These networks are characterized by a highly coherent dynamics \cite{Orlandi2013np,Tibau2013}, and therefore all neurons in the network reduce activity similarly as CNQX is applied. Interestingly, for $[\text{CNQX}]\simeq 400$nM the shown homogeneous culture has almost completely silenced ($\Gamma \simeq -1$), while the clustered culture is still highly active. We repeated this study on $4$ different realizations of each culture type and observed that, on average, the critical concentration $[\text{CNQX}]_\text{C}$ at which activity complete stopped was $1.6 \mu$M for clustered and $0.5 \mu$M for homogeneous networks (Figure~\ref{fig:resilience}C).

Figure~\ref{fig:resilience}D shows the results for the resistance of the networks to node deletion as a consequence of direct photo--damage to the neurons. As can be observed, homogeneous cultures decay in activity much faster than the clustered ones, pinpointing the general resistance of clustered cultures to structural failure.

\section*{Discussion}

Clustered neuronal cultures have a unique self--organizing potential. An initially isotropic ensemble of individual neurons quickly group to one another to constitute a stable configuration of interconnected clusters of tightly packed neurons. The formation of the clustered network is primarily a passive process governed by the pulling forces exerted by the neurites. Interestingly, aggregation occurs even in the absence of glial cells and neuronal activity \cite{Segev2003Electrically}, and is maintained up to the degradation of the culture \cite{Segev2003Electrically,Santos2013,shefi2002morphological}. Our work shows that this self--organizing process drives the network towards specific dynamic states, which shape a topology of the functional network that is distinctively assortative. We note that the number of clusters and their distribution are initially random. Therefore, a wide spectrum of physical circuitries and functional topologies are in principle attainable. However, in all the studied cultures, the network drives itself towards markedly assortative topologies with a `rich--club' core. The emergence of these distinct topological traits, concurrently with a stronger network's resilience to activity deterioration, pictures a self--organizing mechanism that enhances network survival by procuring a robust architecture and dynamic stability.

We remark that the link between assortativity and resilience is based on the comparison between the response of clustered and homogeneous cultures upon the same perturbation. To obtain conclusive evidences that assortativity confers resilience traits exclusively from topology, we would require an experimental protocol in which we could arbitrarily `rewire' the connectivity between clusters, or shape in a control manner different circuitries while preserving the number of nodes in the network. Although these strategies are certainly enlightening, they are of difficult development and a major  experimental challenge.

We infer the functional connectivity maps of the clustered networks from their spontaneous dynamics. We considered small--sized networks to simultaneously access the entire population ($\simeq 30$ clusters). The approach that we have used to characterize this functional connectivity is based on the analysis of the time delays between consecutive clusters' activations. The uniqueness of our approach is to use these time delays to provide a direct measure of causality, giving rise to a functional network that is both directed and weighted, with the weights given by a decaying function that follows the frequency of the delay between pairs of clusters. Our formulation is simple and naturally derives from the intrinsic dynamics of the network.

We used two main parameters to quantitatively construct the directed functional network, namely the cut--off time for causality, and the variance $c$ of the Gaussian--like weighting function. The cut--off time is set to $200$ ms, two times the maximum measured time delay between consecutive activations. The importance of the cut--off is first to discriminate two successive bursting episodes, and second to exclude individual firing events from an actual cascade of activations. Although these individual firings account for less than 2\% of the total activations, they may occur in regions of the culture that are physically distant ---though temporary close--- from an actual sequence, and therefore they would add spurious, long--range functional connections to the network.

On the other hand, the variance $c$ is obtained from a Gaussian fit of the distribution of consecutive activation delays within bursts. The value of $c$ is specific for each culture to take into account  particular differences in the dynamics of the network, specifically the culture days {\em in vitro}  or the number of clusters (Figure S2), parameters that could affect the delay times of activation. Young cultures for instance exhibit longer time delays between pairs of clusters, leading to a distribution $\text{f}(\Delta t)$ shifted towards higher values and therefore a larger $c$.

We tested that the obtained functional networks were stable upon variation of the above parameters. In particular, to examine whether the choice of the cut--off does or does not substantially affect the features of the generated functional network, we performed a sensitivity analysis on this parameter. As the process of generating the network from the sets of bursts is deterministic, we analyzed the influence of the cut--off value on the formed groups of firings. To quantify the variation on the bursts generated for different values of the cut--off, we calculated the variation of information \cite{varinfo} between the grouping of bursts at a certain cut--off value and the previous one as a measure to assess their difference (Figure S3). In the case of clustered cultures, we found that for values of cut--off of $200 \pm 50$ ms the variation of information is, on average, on the order of $10^{-2}$. In the homogeneous case, for cut--off values of $10 \pm 5$ ms, this value is on the order of $10^{-3}$. This means that varying the cut--off values in these regions does not substantially change the grouping of the bursts, and therefore the generated networks are equivalent.

To assess the goodness of our construction in inferring the functional connectivity of the clustered networks, we compared our connectivity maps with those procured by information theoretic measures, such as Mutual Information or Transfer Entropy, applied to the original fluorescence recordings. These approaches have been used to draw the topological properties of neuronal networks {\em in vitro}, both in electrode recordings \cite{Bettencourt2007,Garofalo2009} and calcium fluorescence imaging \cite{Stetter:2012fe,Orlandi:2014-TE}. The comparison of our method with these theoretic measures showed that the identified functional links were fundamentally the same, with small quantitative differences associated to the particular weighting procedures.

Our functional networks consistently maintained high assortativity values, and along a wide range of days {\em in vitro}.  We also observed that, by contrast, the assortativity analysis in homogeneous cultures procured neutral or low assortativity values, a result that is supported by other studies in homogeneous networks similar to ours \cite{Bettencourt2007}. In our study, we have seen that the clustered, assortative networks exhibit a higher resilience of the network to damage compared to the homogeneous, non--assortative ones. Different studies also highlighted the importance of assortativity and the `rich--club' phenomenon on higher--order structures of the network, in particular resilience, hierarchical ordering and specialization \cite{Colizza2006RC,Sporns2011Human}.

Several studies in brain networks advocate that the functional connectivity reflects the underlying structural organization \cite{Deco2010emerging,Honey10022009,Honey12062007}. To shed light on this interrelation in our cultures, we would need the identification of all the physical links between clusters. The top images of Figure~\ref{fig:networks} indeed suggest that some structural connections could be delineated by a simple visual inspection. However, we observed by green fluorescence protein (GFP) transfection that physical connections have long extensions and may easily link several clusters together, and not just in a first--neighbor manner as seen in the images. Since the images provide a very poor subset of the entire structural layout, a complete description of the physical circuitry must be carried out before comparing the structural and functional networks. Such a detailed identification is difficult, and requires the use of a number of connectivity--labeling techniques. Nevertheless, for the connections that we could visualize, we draw two major conclusions. First, that neither the width of the physical connections nor the size of the clusters were related to a particular trait of the functional links, such as the weight of the connections or the strength of the nodes (Figure S4). And, second, that our construction inferred strong functional links between clusters that were not directly connected in a physical manner, highlighting the importance of indirect paths as well as long--range coupling in the flow of activity.

The identification of the full set of structural connections would certainly provide invaluable information to investigate the interplay between structure and function in our networks. In this context, the recent work by Santos-Sierra {\em et al} \cite{Santos2013} is enlightening. They analyzed some major structural connectivity traits in clustered networks similar to ours, and observed that the networks were strongly assortative as well. Assortativity emerged at early stages of development, and was maintained throughout the life of the culture. Hence, in clustered cultures, the combined evidences of this study and ours hints at the existence of assortative properties in both structure and function.

An interesting peculiarity of our experiments is that, in most of the studied clustered cultures, the spontaneous bursting episodes comprised of a small subset of clusters rather than the entire network. This activation in the form of groups or moduli is often referred as {\em conditional activity}. It contrasts with the {\em coherent activity} of homogeneous cultures, where the entire network lights up in a short time window during a bursting episode. Given the acute differences in assortativity between clustered and homogeneous cultures, we hypothesize that the modular dynamics by itself increases or reinforces assortative traits in the functional network.

We finally remark that our neuronal cultures are spatial, i.e. embedded in a physical substrate, which imposes constraints to the layout of connections and, in turn, their assortative characteristics \cite{Santos2013,Schmeltzer2014Percolation}. Spatial networks have caught substantial interest in the last years to understand the restrictions ---or advantages--- that metric correlations impose on the structure and dynamics of complex networks \cite{Barthelemy2011}, in particular brain circuits \cite{bullmore2012economy}. V\'ertes {\em et al} showed that spatial constraints delineate several topological properties of functional brain networks \cite{Vertes2012}, and Orlandi {\em et al} showed that the initiation mechanisms of spontaneous activity are governed by metric correlations inherited by the network during its formation \cite{Orlandi2013np}. Strong spatial constraints in clustered networks can be attained by anchoring the neuronal aggregates in specific locations, for instance through carbon nanotubes \cite{Gabay2005Enginereed}. The comparison of the functional maps of such a {\em forced} organization with our {\em free} one is enlightening, and would shed light on the importance of structural constraints in shaping functional connectivity.

To conclude, we have presented a simple yet powerful construction to draw the directed functional connectivity in clustered neuronal cultures. The construed networks present assortativity and `rich--club' features, which are present concurrently with resilience traits. Our analysis has been based on spontaneous activity data, and may certainly vary from evoked activity. Hence, the combined experimental setup and functional construction can be viewed as a model system for complex networks studies, specially to understand the interplay between structure and function, and the emergence of key topological traits from network dynamics. Also, the spatial nature of our experiments may also procure invaluable data to understanding the role of short-- and long--range connections in network dynamics; or to investigate the targeted deletion of the high degree nodes that shape the backbone of the network. The latter is a powerful concept that may assist in a detailed exploration of resilience in neuronal circuits, for instance to model the circuitry--activity interrelation in neurological pathologies.

\section*{Materials and Methods}

\subsection*{Ethics Statement}

All procedures were approved by the Ethical Committee for Animal Experimentation of the University of Barcelona, under order DMAH-5461.

\subsection*{Clustered neuronal cultures}

In our experiments we used cortical neurons from $18-19$ day old Sprague--Dawley rat embryos. Following standard procedures \cite{Segal1992,Orlandi2013np} dissection was carried out in ice--cold L--15 medium enriched with 0.6\% glucose and gentamycin (Sigma-Aldrich). Cortices were gently extracted and dissociated by repeated pipetting.

Cortical neurons were plated onto $13$ mm glass coverslips (Marienfeld-Superior) that incorporated a poly-dimethylsiloxane (PDMS) mold. The PDMS restricted neuronal growth to isolated, circular cavities $3$ mm in diameter. Prior plating, glasses were washed in 70\% nitric acid for 2 h, rinsed with double--distilled water (DDW), sonicated in ethanol and flamed. In parallel to glass cleaning, and following the procedure described by Orlandi \emph{et al.}~\cite{Orlandi2013np}, several $13$ mm diameter layers of PDMS $1-2$ mm thick were prepared and subsequently pierced with $3$ mm diameter biopsy punchers (Integra-Miltex). Each pierced PDMS mold typically contained $4$ to $6$ cavities.  The PDMS molds were then attached to the glasses and the combined structure autoclaved at $120^{\circ}$C, firmly adhering to one another. For each dissection we prepared $12$ identical glass-PDMS structures, giving rise to about $80$ cultures of $3$ mm in diameter. Neurons were plated in the PDMS cavities with a nominal density of $500$ neurons/mm$^2$, and incubated in plating medium at $37^{\circ}$C, $5\%$ CO$_{2}$ and $95\%$ humidity. Plating medium consisted in $5\%$ of foetal calf serum (FCS, Invitrogen), $5\%$ of horse serum (HS, Inivtrogen), and $0.1$\% B27 (Sigma) in MEM Eagle's-L-glutamate (Invitrogen). MEM was enriched with gentamicin (Sigma), the neuronal activity promoter Glutamax (Sigma) and glucose.

Upon plating, the absence of adhesive proteins in the glass substrate rapidly favored cell--cell attachment and, gradually, the formation of islands of highly compact neuronal assemblies or \emph{clusters} that minimized the surface contact with the substrate. Clustered cultures formed quickly. By \emph{day in vitro} (DIV) 2 the culture encompasses dozens of small aggregates that coalesce and grow in size as the culture matures. Spontaneous activity and connections between clusters were observed by DIV $4-5$. Clusters at this stage of development also anchored at the surface of the glass and, although they continued growing and developing connections, their number and position remained stable along the next $2$ weeks. At the moment of measuring, each PDMS cavity contained an independent culture formed by $20-40$ interconnected clusters.

Clustered cultures were maintained for about $3$ weeks, as follows. At DIV $5$ the medium was switched from plating to \emph{changing medium} (containing $0.5\%$ FUDR, $0.5\%$ Uridine, and $10\%$ HS in enriched MEM) to limit glial cell division. Three days later, the medium was replaced to \emph{final medium} (enriched MEM with $10\%$ HS), which was then refreshed periodically every three days.

\subsection*{Homogeneous neuronal cultures}

Overnight exposure of the glass coverslips to poly-l-lisine (PLL, Sigma) provided a layer of adhesive proteins for the neurons to quickly anchor upon seeding, leading to cultures with a homogeneous distribution of neurons over the substrate. The remaining steps in the preparation and maintenance of the cultures were identical as the clustered ones, i.e.\ we used the same nominal neuronal density for plating, we included PDMS pierced molds to confine neuronal growth in cavities $3$ mm in diameter, and we refreshed the culture mediums in the same manner.

\subsection*{Experimental setup and procedure}

\subsubsection*{Standard experiments}
To measure the spontaneous activity in the clustered networks we used cultures at day {\it in vitro} (DIV) $5-16$, i.e.\ covering about two weeks of development. Cultures started to degrade by DIV $25$, and therefore we did not use cultures older than $3$ weeks in our experiments.

Activity in neuronal cultures was monitored through fluorescence calcium imaging \cite{Takahashi2010,Grienberger2012-Imaging}, which allows the detection of neuronal activity by the binding of Ca$^{+2}$ ions to a fluorescence probe upon firing.
Prior to recording, the cultures under study were incubated for 40 min in External Medium (EM, consisting of $128$ mM NaCl, $1$ mM CaCl$_2$, $1$ mM MgCl$_2$, $45$ mM sucrose,
$10$ mM glucose, and $0.01$ M Hepes; pH 7.4) in the presence of Fluo-4-AM (Invtrogen). We used $4$ $\mu$l Fluo4 in a volume of 2 ml EM. We incubated a glass coverslip containing $4$ cultures within the PDMS cavities at once, allowing for the simultaneous recording of different cultures or the selection of cultures with specific traits. After incubation, the cultures were washed with fresh EM and placed in the observation chamber, consisting of a standard glass bottom culture dish, filled with 4 ml EM, and with its wall and cover screened from external light. To minimize accidental damage to the aggregates during the manipulation of the cultures, the PDMS pierced mold was left in contact with the glass during both incubation and the actual experiment.

The observation chamber was mounted on Zeiss Axiovert inverted microscope equipped with a high--speed CMOS camera (Hamamatsu Orca Flash 2.8). We used an objective of $2.5$X combined with a $0.32$X optical zoom. These settings provided a final field of view of $7.6\times3.4$ ($\text{width}\times \text{height}$) mm$^2$ that supported the recording of $1$ or $2$ PDMS--confined cultures simultaneously. Individual frames were acquired as 8--bit grey--scale images, a size of $940\times400$ pixels, and a spatial resolution of $8.51$ $\mu$m/pixel. All experiments were carried out at room temperature.

The fluorescence signal of the clusters' spontaneous activity was recorded with the software Hokawo 2.5, provided by the camera vendor (Hamamatsu Photonics). We used acquisition speeds in the range $83-100$ frames per second (fps), corresponding to a time interval of $12-10$ ms between consecutive frames. These acquisition speeds were selected to optimize the balance between image quality, sufficient time resolution, and minimum light intensity. The latter was particularly important to minimize photo-damage and photo-bleaching, and allowed neuronal cultures to be studied with optimal conditions for at least 3 h. However, the combination of high acquisition speeds and high resolution images resulted in large data files ---of at least 150 GB per hour of recording--- that had to be stored and analyzed. We therefore limited most of our experiments to about 1 h of recording, which was sufficient to reliably build the functional networks, as described in the `Control experiments' section.

Measurements in homogeneous cultures were carried out in the same way, with the only difference that the recording speed was increased to 100 -- 150 fps to take into account the fast propagation of activity fronts in these preparations, as observed for instance in the study of Orlandi \emph{et al.} \cite{Orlandi2013np}.

\subsubsection*{Resilience experiments}

We considered two groups of resilience experiments. In a first group, we monitored the gradual degradation of network activity due to photo-damage. In a second group, we measured the decay in activity as a consequence of the gradual weakening of the excitatory connectivity.

In the first group of measurements, we first considered a clustered culture and measured its spontaneous activity uninterruptedly along 2 hours, with neurons continuously exposed to a light radiation $4$ times stronger than normal. We then divided the sequence in blocks of 30 min, and determined, for each block, the average network activity by counting the number of bursting episodes within the block. Next, we switched to a homogeneous culture from the same batch (i.e. identical nominal density and age) and carried out the same protocol. In total we carried out $6$ measurements for each kind of culture, and finally analyzed the decay in activity as a function of time.  Although the radiation over the culture was homogeneous, the actual area occupied by the neurons was different between homogeneous and clustered cultures. Neurons in homogeneous cultures formed a monolayer that covered $(34 \pm 4)\%$ of the available area, corresponding to about $2.4$ mm$^2$ for the $3$ mm diameter wells. In clustered cultures, neurons were tightly packed in slightly three--dimensional structures, giving rise to a lower spatial coverage of $(26 \pm 5)\%$, i.e. about $1.8$ mm$^2$. Hence, homogeneous cultures were effectively more exposed to light than their clustered counterparts. To take this spatial variability into account, the `total radiation' received by the neurons in a given experiment was quantified as the duration of the light exposure times the area occupied by the neurons in the studied culture. We then averaged the results over the $6$ different culture realizations of each type, and binned nearby values for clarity. The comparison of the activity--radiation plots between the two networks (Figure~\ref{fig:resilience}D) indicated which topology exhibited higher resistance to degradation in neuronal activity.

In the second group of measurements, we compared the change in activity between a clustered and a homogeneous culture during gradual weakening of neuronal connectivity. The weakening was achieved by progressive application of CNQX \cite{Soriano2008-Development,Tibau2013}, an AMPA-glutamate receptor antagonist in excitatory neurons (see also `Pharmacology'). We first measured the clustered network and thereafter the homogeneous one. In both cases, we first recorded spontaneous activity at $[\text{CNQX}]=0$ nM, and used the average firing rate $\gamma_0$ as reference for the subsequent steps. We then increased the concentration of the drug to a preset value, waited $5$ min for the drug to take effect and measured again for $15$, computing the new average firing rate $\gamma$. We switched to a second preset values, and repeated the procedure until activity ceased. The relative decay in activity $\Gamma \equiv (\gamma - \gamma_0)/\gamma_0$ is used to illustrate the gradual fall of activity in Figure~\ref{fig:resilience}C. The critical concentration [CNQX]$_\text{C}$ at which activity is absent along the $15$ min of recording ($\Gamma =-1$) hints at the robustness of network dynamics to a global failure of its connectivity.

\subsection*{Pharmacology}

The pharmacological protocols described below were used identically in clustered and homogeneous cultures.

\subsubsection*{Inhibitory connections}
The \emph{in vitro} networks contain both excitatory and inhibitory connections. However, for sake of simplicity in the comparison between experiments and model, in the experiments at DIV $6$ and above we completely blocked $\gamma$-aminobutyric acid (GABA) inhibitory synapses with $40$ $\mu$M of the antagonist bicuculine methiodide (Sigma). The drug was applied $5$ min before the actual recordings for the drug to take effect. Spontaneous activity in our experiments is therefore solely driven by excitatory connections. Although the balance between excitation and inhibition shapes the major traits of spontaneous activity \cite{Orlandi2013np,Tibau2013}, such as the average firing rate of the network, we verified that the presence of inhibition did not qualitatively modify the results presented here.

We left active inhibitory synapses for experiments at DIV $5$ since at this early stages of development GABA has a depolarizing effect and therefore an excitatory action \cite{Ganguly2001-GABA,Soriano2008-Development,Tibau2013}. Its blockade would effectively reduce excitation and silence the network.

\subsubsection*{Network connectivity weakening through CNQX}

In the studies of network resilience to the weakening of connectivity, we studied the decay in spontaneous activity as a result of the gradual application of 6-cyano-7-nitroquinoxaline-2,3-dione (CNQX, Sigma), an AMPA-glutamate receptor antagonists in excitatory neurons. For $\text{[CNQX]}=0$ the connectivity strength between neurons is maximum. As [CNQX] is administered, the efficacy of excitatory connections steadily diminishes, which is accompanied by a reduction in spontaneous activity (see e.g.\cite{Tibau2013} for illustrative data). High CNQX concentrations lead to a complete halt in activity. In the measurements we used CNQX concentrations in the range $0-2000$ nM, in quasi-logarithmic steps. We left the culture unperturbed for 5 min after each CNQX application for the drug to reach steady--state effects.

\subsection*{Data analysis for clustered cultures}

The acquired images (recorded at a typical speed of $100$ fps) were first analyzed with the Hokawo 2.5 software to extract the fluorescence intensity of each cluster as a function of time. The regions of interest (ROIs) were chosen manually and typically covered an area of $40\times40$ pixels, each ROI corresponding to a single cluster. As illustrated in Figure~\ref{fig:experiments}C and Figure~\ref{fig:experiments}F, activity is characterized by a stable baseline (resting state) interrupted by peaks of fluorescence that correspond to clusters' firings. At the onset of firing, the fluorescence signal increases abruptly due to the fast intake of Ca$^{2+}$ ions. Fluorescence then reaches a maximum, and slowly decays back to the baseline in $2-5$ s.

The algorithm that we used to detect the onset of firing for each cluster was as follows. We first corrected the fluorescence signal $\tilde{F}(t)$ from small drifts, and calculated the resting fluorescence level $F_0$ by discarding the data points with an amplitude two times above the standard deviation (SD) of the signal. The corrected signal was then expressed as $F(t) \equiv \Delta \tilde{F} / F_0 = (\tilde{F}-F_0) / F_0$. We next took $F(t)$ and computed its derivative $\dot{F}(t)$ in order to detect fast changes in the fluorescence signal. Finally, the onset of ignition in cluster was defined as the time where a maximum in $\dot{F}(t)$ was accompanied by values of $F(t)$ two times above the SD of the background signal, and for at least 5 frames.

\subsubsection*{Reliability in detecting the clusters' ignition times}

Three major tests were carried out to assess the reliability of our analysis. In a first one, we measured spontaneous activity at $200$ fps, i.e. twice the standard recording speed, but used stronger light to compensate for the lower exposure time. We next analyzed the data, re-sampled the image sequence down to $100$ fps and compared the results with the original acquisition. We observed that the detection of the onset times improved only by about 15\%, which did not justify the excess of light and the associated damage to the neurons.

In a second test, we measured spontaneous activity in a culture using identical light settings but considering different acquisition rates, namely $100$, $150$, and $200$ fps. We then selected ignition sequences that were as similar as possible in all three measurements, and compared the results. We observed that only in the few cases where the clusters fired with strong amplitudes the increased speed enhanced detection, and again by 15\%. For the rest of the cases, the higher speeds actually worsened the analysis due to the poorer signal-to-noise ratio.

Finally, in a third test, we used sub--frame resolution analysis tools to evaluate the importance of finer ignition times. Following Orlandi \emph{et al.} \cite{Orlandi2013np}, we considered the approach of fitting two straight lines at the vicinity of each initially detected firing. A first fit included the $100$ points of the background signal that preceded ignition, and a second one extended to the $10$ points that correspond to the fast rise in fluorescence. The crossing value of the two lines provided an onset time that refined the initially measured value. The better accuracy effectively increased the discrimination of sequences that were initially identified as simultaneous events. However, since these events are rare (by 5\%), the finer temporal resolution had practically no effect in the construction of the functional networks and the derived analysis.

\subsubsection*{Activity propagation times}

The time delay $t_p$ in the propagation of activity between two connected clusters was measured in control experiments with high acquisition rates. We concluded that $t_p$ varied in the range $10 \leq t_p \leq 100$ ms, with an average value $\bar{t}_p =50$ ms. Other studies in clustered networks provided similar results \cite{Tsai2008Robustness}. With the detection algorithm described above and standard experiments at $83-100$ fps, we could appraise the activation sequence in 93\% of the cases. The remaining 7\% corresponded to clusters that ignited in the same frame or time bin, and were treated as simultaneous events.

\subsection*{Data analysis for homogeneous cultures}

Recordings in homogeneous cultures provided the activity of $\simeq1000$ neurons in an circular area 3 mm in diameter. Neurons were marked individually as regions of interest in the images and the corresponding fluorescence time traces extracted using custom--made software. Ignition times for each neuron were next obtained by using the sub--frame resolution method described above (detailed in Ref. \cite{Orlandi2013np}), and that consisted in fitting two straight lines to the fluorescence data, a first fit encompassing the $100$ points in the background region prior to firing, and a second fit including the $10$ points during the fast rise in fluorescence that follows ignition. The crossing point of the two lines provided the onset of firing.

The extension of this analysis to all the active neurons within a burst, and along all the bursts, finally provided the entire set of ignition sequences. The construction of the directed functional networks for the homogeneous cultures was then carried out identically as the clustered ones.

\subsection*{Additional control experiments}

Recordings in clustered cultures typically lasted for $1$ h and contained between $\simeq 50$ bursts in the quietest networks and $\simeq 450$ bursts in the most active ones. To test whether $50$ bursts sufficed to draw the functional networks, we carried out a control experiment in which we monitored spontaneous activity along $2$ h in a standard clustered culture, measured at DIV $12$ and containing $42$ nodes (Figure S6). We then analyzed the data using two different procedures. In the first one we drew the functional connectivity using the data extracted from the entire recording, and determined its assortativity values. In the second procedure, we separated the recorded sequence in three blocks, each $40$ min long, built the functional connectivity for each block, and computed the respective assortative values. The studied culture fired in a sustained manner at a rate of $1.12$ bursts/min, and procured a total of 134 bursts. Thus, each block typically contained about $45$ bursts.

The results (Figure S6) led to two major conclusions. First, that the functional connectivity is very similar among the blocks, and between any of the blocks and the entire recording, providing assortativity values that are compatible within statistical error. And second, that the first 40 min of recording (with 45 bursts only) sufficed to shape the major traits of the functional network, therefore validating our strategy of using $1$h of acquisition to procure a reliable estimate of the functional connectivity of the network and its assortative traits.

\subsection*{Network assortativity and rich--club}

Here we describe the calculation of the assortativity coefficients, assortativity errors and the rich--club distributions. In the process, we have to define the assortativity for directed weighted networks.

\subsubsection*{Generalization of assortativity to directed weighted networks}

Newman \cite{newman2002assortative} defined {\em assortativity} $\supe{\rho}{P}$ as the Pearson correlation between the degrees of every pair $E$ of linked nodes in the network. More precisely, in the case of directed networks, if $\wout{k}{i}=\sum_j a_{ij}$ is the output degree from node $i$, $\win{k}{j}=\sum_i a_{ij}$ the input degree to node $j$, and $E$ scans all the edges in the network, then
\begin{equation}
  \supe{\rho}{P} =
    \frac{
      \ds\frac{1}{L}\sum_{(i,j)\in E} \left(\wout{k}{i} - \subi{\avg{\wout{k}{}}}{E}\right)
                                      \left(\win{k}{j}  - \subi{\avg{\win{k}{} }}{E}\right)
    }{
      \sqrt{\ds\frac{1}{L}\sum_{(i,j)\in E} \left(\wout{k}{i} - \subi{\avg{\wout{k}{}}}{E}\right)^2}
      \sqrt{\ds\frac{1}{L}\sum_{(i,j)\in E} \left(\win{k}{j}  - \subi{\avg{\win{k}{} }}{E}\right)^2}
    },
\end{equation}
where
\begin{equation}
  \subi{\avg{\wout{k}{}}}{E}
    = \frac{1}{L}\sum_{(i,j)\in E} \wout{k}{i}
    = \frac{1}{L}\sum_{i=1}^N\sum_{j=1}^N a_{ij} \wout{k}{i}
    = \frac{1}{L}\sum_{i=1}^N (\wout{k}{i})^2,
\end{equation}
\begin{equation}
  \subi{\avg{\win{k}{}}}{E}
    = \frac{1}{L}\sum_{(i,j)\in E} \win{k}{j}
    = \frac{1}{L}\sum_{i=1}^N\sum_{j=1}^N a_{ij} \win{k}{j}
    = \frac{1}{L}\sum_{j=1}^N (\win{k}{j})^2,
\end{equation}
\begin{equation}
  L = \sum_{(i,j)\in E} a_{ij} = \sum_{i=1}^N\sum_{j=1}^N a_{ij}.
\end{equation}
After some algebra, assortativity may also be written as
\begin{eqnarray}
  \supe{\rho}{P} & = &
    \frac{
      \ds\sum_{i,j} a_{ij}\left(\wout{k}{i} - \subi{\avg{\wout{k}{}}}{E}\right)
                          \left(\win{k}{j}  - \subi{\avg{\win{k}{} }}{E}\right)
    }{
      \sqrt{\ds\sum_{i,j} a_{ij}\left(\wout{k}{i} - \subi{\avg{\wout{k}{}}}{E}\right)^2}
      \sqrt{\ds\sum_{i,j} a_{ij}\left(\win{k}{j}  - \subi{\avg{\win{k}{} }}{E}\right)^2}
    }
  \\
    & = &
    \frac{
      \ds\sum_{i,j} a_{ij}\left(\wout{k}{i} - \subi{\avg{\wout{k}{}}}{E}\right)
                          \left(\win{k}{j}  - \subi{\avg{\win{k}{} }}{E}\right)
    }{
      \sqrt{\ds\sum_{i} \wout{k}{i}\left(\wout{k}{i} - \subi{\avg{\wout{k}{}}}{E}\right)^2}
      \sqrt{\ds\sum_{j} \win{k}{j} \left(\win{k}{j}  - \subi{\avg{\win{k}{} }}{E}\right)^2}
    }
\end{eqnarray}

This definition of assortativity is applicable to all kind of networks, either undirected, directed, weighted or unweighted. For weighted networks as in our case, the strength of the nodes carries important information about the structure of the network, and thus it would be useful to know the correlation between the strengths instead of the degrees. Since in this case each edge carries a weight, it seems logical that edges with higher weight should have a larger contribution to the correlation. Therefore, we define the {\em weighted assortativity} $\supe{\rho}{PW}$ as the Pearson weighted correlation between the strengths of the nodes. In mathematical terms, if $w_{ij}$ is the weight of the link from node $i$ to node $j$ (zero if there is no link), then $\wout{s}{i}=\sum_j w_{ij}$ and $\win{s}{j}=\sum_i w_{ij}$ are the output and input strengths, then
\begin{equation}
  \supe{\rho}{PW} =
    \frac{
      \ds\sum_{i,j} w_{ij}\left(\wout{s}{i} - \subi{\avg{\wout{s}{}}}{E}\right)
                          \left(\win{s}{j}  - \subi{\avg{\win{s}{} }}{E}\right)
    }{
      \sqrt{\ds\sum_{i} \wout{s}{i}\left(\wout{s}{i} - \subi{\avg{\wout{s}{}}}{E}\right)^2}
      \sqrt{\ds\sum_{j} \win{s}{j} \left(\win{s}{j}  - \subi{\avg{\win{s}{} }}{E}\right)^2}
    },
\end{equation}
where
\begin{equation}
  \subi{\avg{\wout{s}{}}}{E}
    = \frac{1}{S}\sum_{(i,j)\in E} w_{ij}\wout{s}{i}
    = \frac{1}{S}\sum_{i} (\wout{s}{i})^2,
\end{equation}
\begin{equation}
  \subi{\avg{\win{s}{}}}{E}
    = \frac{1}{S}\sum_{(i,j)\in E} w_{ij}\win{s}{j}
    = \frac{1}{S}\sum_{j} (\win{s}{j})^2,
\end{equation}
with $S$ the total strength of the network, i.e.
\begin{equation}
  S = \sum_{(i,j)\in E} w_{ij} = \sum_{i,j} w_{ij}.
\end{equation}

Litvak {\em et al.} \cite{spearman2013} showed that in disassortative networks the magnitude of the standard assortativity decreases with network size, a problem that was solved by replacing the Pearson correlation $\supe{\rho}{P}$ with the Spearman correlation, thus obtaining a {\em Spearman assortativity} $\supe{\rho}{S}$. Spearman rank correlation is calculated in the same way that the Pearson correlation but substituting the values (in this case, the degrees of the nodes) by their respective ranks, i.e.\ their position when the values are sorted in ascending order. This leads us to define the {\em Spearman weighted assortativity} $\supe{\rho}{SW}$ as the Spearman weighted correlation between the strengths of the nodes at both ends of each edge in the network.

The estimation of the error in the assortativity value (for any of the previous four variants) can be computed in several ways, for instance through the jackknife method, the bootstrap algorithm, or by using the Fisher transformation \cite{newman2003mixing,Efron1979-SIAM-Computers}. We used in our work the bootstrap algorithm, and considered $1000$ random samples of the data.

\subsubsection*{Rich--club analysis}

The rich--club analysis computes the degree--degree (or weight--weight) correlation distributions, and respect to a null case of non--correlated degrees (or weights). It allows us to reinforce the assortativity analysis presented before. Assortative mixing, in principle, will induce a rich--club effect that should be clearly detectable for a wide range of degrees (or weights). We will first introduce the formulation for the calculation of rich--club in weighted networks as presented in \cite{mangels_rich}, and afterwards we will extend it to the case of weighted directed networks.

The rich--club score is calculated as follows:
\begin{equation}
  \phi^{\text{unc}}_{s_T}=\frac{W_{s_T}}{W^{\text{unc}}_{s_T}},
  \label{eq:rich}
\end{equation}
where $\phi_{s_T}$ is the rich--club score relative to the uncorrelated null case, $W_{s_T}$ is the sum of the weights of the links of the subgraph formed only by those nodes whose strengths are higher than $s_T$,
\begin{equation}
  W_{s_T}=\sum_{i \in v_{s_T}} \sum_{j \in v_{s_T}} w_{ij},
  \label{eq:rich_num}
\end{equation}
and $W^{\text{unc}}_{s_T}$ is the corresponding value in the case of uncorrelated strengths,
\begin{equation}
  W^{\text{unc}}_{s_T}= \avg{s} \frac{\ds\sum_{i \in v_{s_T}} \sum_{j \in v_{s_T}}s_i s_j}{\ds N\avg{s}^2-\avg{s^2}}.
  \label{eq:rich_den}
\end{equation}
The term $v_{s_T}$ designates the subset of nodes $i$ such that $s_i>s_T$. $N$ is the total number of nodes in the network, and $\langle s \rangle$ and $\langle s^2 \rangle$ are the first and second moments of the strength distribution. The ratio $\phi_{s_T}$ is calculated for all values of $s_T$, and ranges from the minimum value of strength in the network to the maximum.
This ratio indicates the presence or absence of a rich--club in the network: a network shows a rich--club effect when the high values of $s_T$ give a ratio above 1.

To calculate this ratio on our functional networks, we have to consider that the network is not only weighted but also directed. Therefore, we need to adapt the former formulation for the case of weighted directed networks. For this reason we will consider the in-- and out--strength of each node, expressed as $\win{s}{i}$ and $\wout{s}{i}$ respectively. In the directed formulation, Eqs.~(\ref{eq:rich}) and~(\ref{eq:rich_num}) remain unchanged. However, $v_{s_T}$ must be redefined as
\begin{equation}
  v_{s_T}=\{ i \,| \,\win{s}{i}+\wout{s}{i} > 2s_T\},
\end{equation}
and the term $W^{\text{unc}}_{s_T}$ becomes
\begin{equation}
  W^{\text{unc}}_{s_T}= \avg{s} \frac{\ds\sum_{i \in v_{s_T}} \sum_{j \in v_{s_T}} \wout{s}{i}\win{s}{j}}{\ds N\avg{s}^2-\avg{\wout{s}{}\win{s}{}}},
\end{equation}
where the averages are calculated as
\begin{equation}
  \avg{s} = \frac{1}{N} \sum_i \wout{s}{i} = \frac{1}{N} \sum_i \win{s}{i}, \ \ \
  \avg{\wout{s}{}\win{s}{}} = \frac{1}{N} \sum_i \wout{s}{i} \win{s}{i}.
\end{equation}
This formulation allows us to calculate the rich--club coefficient for weighted directed networks. Note that for an undirected network (where $\win{s}{i} = \wout{s}{i}$) the latter formulation reduces to the original one.

\subsection*{Alternative construction of the functional network based on mutual information}

Mutual information\cite{Garofalo2009,Singh2010} is a particular case of the Kullback-Leibler divergence \cite{kullback1951information}, an information-theoretic measure of the distance between two probability distributions. In fact, the mutual information between two stochastic variables $X$ and $Y$ provides an estimation of the amount of information gained about $X$ when $Y$ is known.

Let us indicate by $\{s_{\ell}^{(i)}\}$ the time series corresponding to the $i$-th cluster, with $\ell=1,2,...,L$ and $L$ the total number of time frames involved in the observation process. The time series adopted for the successive analysis are obtained by mapping the observed train of cluster activations to another time series termed \emph{walk}, defined by
\begin{eqnarray}
\label{eq:walk}
x_{\ell}^{(i)} = \sum_{l=1}^{\ell}\left[ s_{l}^{(i)}-\langle s_{\ell}^{(i)}\rangle \right].
\end{eqnarray}

In the specific case of our analysis, the mutual information between two time series $\{x_{\ell}^{(i)}\}$ and $\{x_{\ell}^{(j)}\}$, corresponding to two different clusters, is interpreted as the amount of correlation between the dynamics of cluster $i$ and $j$. In general, the time scale of the correlation between two time series is not known \emph{a priori}. Such a time scale corresponds to the time delay required to maximize the gain of information. Therefore, in the spirit of Fraser and Swinney \cite{fraser1986independent}, we define the time delayed mutual cross information between $\{x_{\ell}^{(i)}\}$ and $\{x_{\ell}^{(j)}\}$ by
\begin{eqnarray}
\label{eq:mutual}
I(x^{(i)},x^{(j)};\tau)=-\sum_{\mu,\nu}p_{\mu\nu}^{(i,j)}(\tau)\log\frac{p_{\mu\nu}^{(i,j)}(\tau)}{p_{\mu}^{(i)}p_{\nu}^{(j)}},
\end{eqnarray}
where $\mu$ and $\nu$ are indices running over some partition of the observed time series. In Eq.\,(\ref{eq:mutual}), $p_{\mu}^{(i)}$ indicates the probability to find a value of time series $\{x_{\ell}^{(i)}\}$ in the $\mu$-th interval, $p_{\nu}^{(j)}$ is the probability to find a value of time series $\{x_{\ell}^{(j)}\}$ in the $\nu$-th interval, whereas $p_{\mu\nu}^{(i,j)}$ denotes the joint probability to observe a firing from the $i$-th cluster falling in the $\mu$-th interval and a firing from the $j$-th cluster falling in the $\nu$-th interval exactly $\tau$ time frames later.

For the sake of simplicity, in the following we will adopt the more concise notation $I_{ij}(\tau)=I(x^{(i)},x^{(j)};\tau)$ to indicate the time delayed mutual cross information. Finally, in order to gain the highest amount of information about the dynamics of cluster $i$ by observing cluster $j$, we consider only the maximum value $I_{ij}^{\max}=\max_{\tau}[I_{ij}(\tau)]$ of $I_{ij}(\tau)$ with respect to the time delay $\tau$.

We estimate the importance of the observed amount of correlation by performing the above analysis on surrogate data. Surrogates adopted in this study are time series generated by randomly reshuffling the temporal observations of the firing series $\{s_{\ell}^{(i)}\}$, for each cluster separately. Such a procedure destroys any correlation between pairs of time series while preserving the empirical probability distribution, thus allowing to test the null hypothesis that the observed correlation is obtained by chance.

We indicate by $\{\tilde{x}_{\ell}^{(i)}\}$ the walk corresponding to the surrogate obtained from time series $\{x_{\ell}^{(i)}\}$ and with $\tilde{I}_{ij}(\tau)$ the time delayed mutual cross information between $\{\tilde{x}_{\ell}^{(i)}\}$ and $\{\tilde{x}_{\ell}^{(j)}\}$. We perform 200 independent random realizations of surrogates for each pair $(i,j)$ and we estimate the  corresponding expected value $\langle \tilde{I}_{ij}^{\max}\rangle$ of the maximum mutual cross--information, as well as the root mean square $\tilde{\sigma}_{ij}$ of the underlying distribution.

Hence, we fix \emph{a priori} the significance $\alpha$ of the hypothesis testing and we estimate the $z$-score corresponding to each pair $(i,j)$ by $z_{ij}=(I_{ij}^{\max}-\langle \tilde{I}_{ij}^{\max}\rangle)/\tilde{\sigma}_{ij}$. Therefore, the observed correlation between cluster $i$ and $j$ is said to be statistically significant if $1-\erf(z_{ij}/\sqrt{2})\leq \alpha$, where $\erf$ is the standard error function.
Finally, we obtain the functional network of clusters by building the weight matrix $\mathbf{W}$ whose elements are defined by $w_{ij}=z_{ij}$ if $1 - \erf(z_{ij}/\sqrt{2}) \leq \alpha$, and $w_{ij}=0$ if $1 - \erf(z_{ij}/\sqrt{2}) > \alpha$.

\section*{Acknowledgments}

The authors acknowledge S. R\"udiger and J. G. Orlandi for fruitful discussions and insight, and to E. Tibau for technical assistance. Research was supported by the Ministerio de Ciencia e Innovaci\'on (Spain) under projects FIS2010-21924-C02-02, FIS2011-28820-C02-01 and FIS2012-38266-C02-01. We also acknowledge the Generalitat de Catalunya under projects 2009-SGR-00014 and 2009-SGR-838, and the EU FET (MULTIPLEX 317532). A.A. also acknowledges partial financial support from the ICREA Academia and the James S.\ McDonnell Foundation.



\newpage

\section*{Supporting Information Figures}

\begin{figure}[!t]
\begin{center}
\includegraphics[width=12cm]{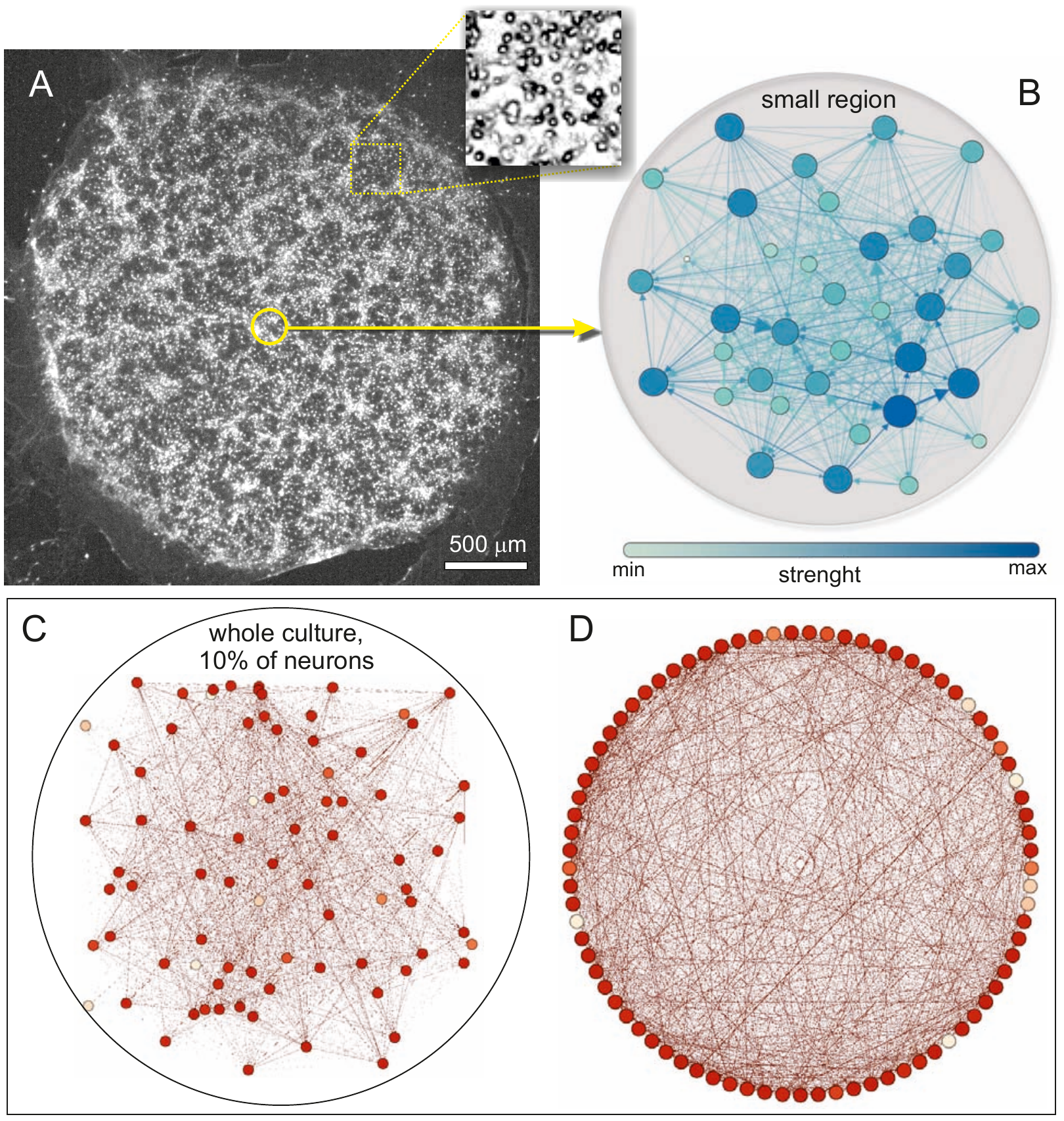}
\end{center}
\caption{
\textbf{Figure S1: Homogeneous cultures.} \textbf{A} The seeding of neurons
in cover glasses previously coated with poly-l-lysine gives
rise to neuronal cultures with a quasi--homogeneous
distribution of neurons. A typical circular culture $3$ mm in
diameter contains about $2000-3000$ neurons that can be well
identified either as bright spots in the fluorescence
recordings or as circular objects in bright--field images
(small panel on the right). Spontaneous activity in these
homogeneous cultures is typically recorded at 100 frames/s,
which suffices to extract the time delays between consecutive
neuronal activations. The particular experiment shown here
corresponds to network `P' of the main text, with a total of
814 neurons manually selected over the images and monitored
along 45 min. The analysis of their spontaneous activity traces
is analyzed in the context of our model, finally procuring the
functional connectivity network and its topological properties.
\textbf{B} Given the large number of nodes analyzed and the high
average degree of the resulting network ($\sim550$ functional
connections per neuron), a representation of the complete
functional network is unpractical. As an example of the
obtained functional networks, we here show all the functional
links within a small region placed in the center containing 40
neurons. Connections are both color and thickness coded
according to their weight. Nodes are color coded according to
their strength. \textbf{C} As an alternative representation, we show
here a 10\% of the population (81 neurons randomly chosen),
each neuron showing the 10\% of its links. Nodes and links are color
coded according to their strength and weight, respectively. \textbf{D} A ring graph of the same neurons shows that most of them display a similar connectivity, in
contrast to the strong modularity and variability in
connectivity exhibited by the clustered cultures.
}\end{figure}
\newpage

\begin{figure}[!t]
\begin{center}
\includegraphics[width=16cm]{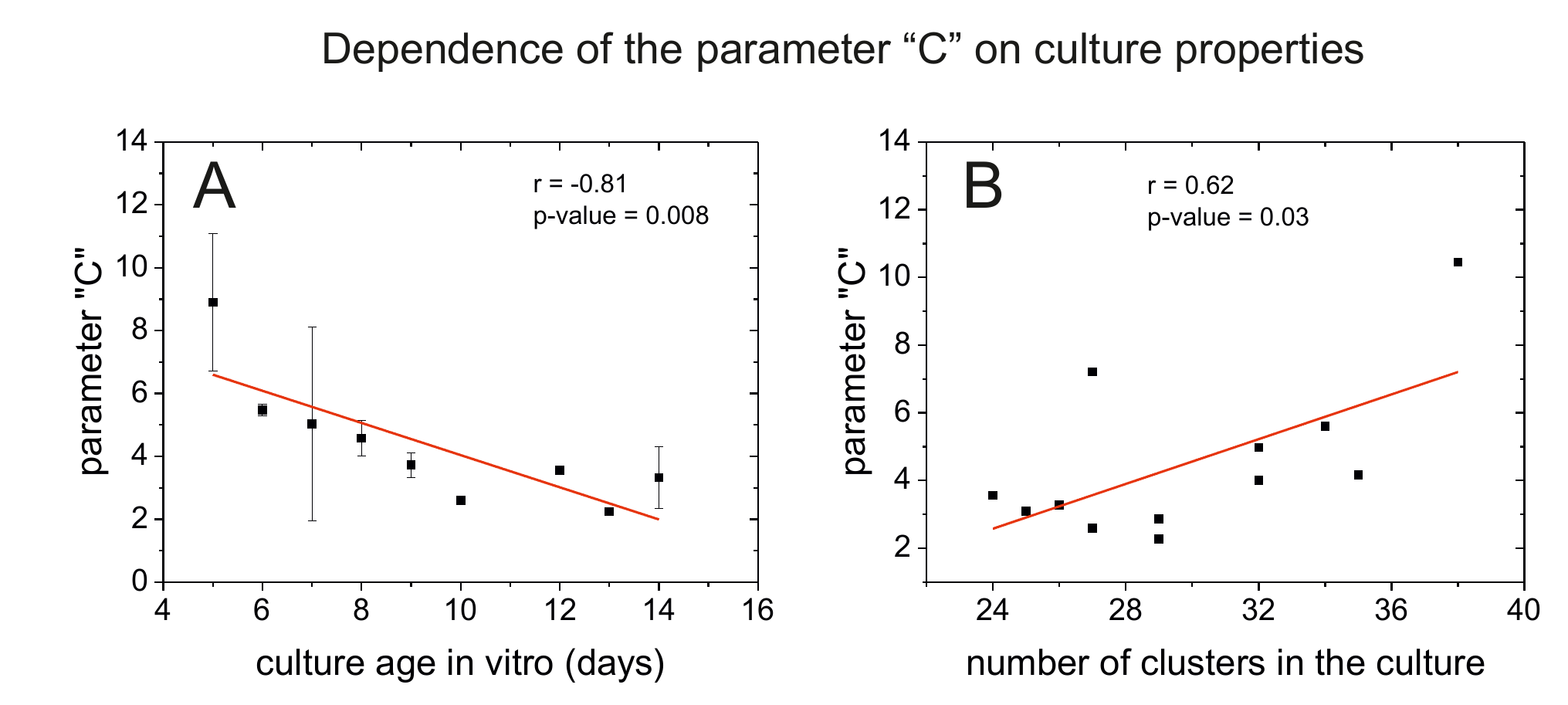}
\end{center}
\caption{
\textbf{Figure S2: Variance $c$ and culture properties.}
\textbf{A} The variance $c$ is obtained from the Gaussian fit of the activation delays between pairs of clusters. The plot shows that $c$ decreases with the culture age {\em in vitro}, indicating that young cultures display a slower dynamics (larger delay times and therefore larger variance) than mature cultures. \textbf{B} The variance $c$ increases with the number of clusters in the culture, indicating a broader and richer distribution of time delays as more clusters participate in the dynamics of the network. Errors bars in \textbf{A} show standard deviation.
}\end{figure}
\newpage

\begin{figure}[!t]
\begin{center}
\includegraphics[width=16cm]{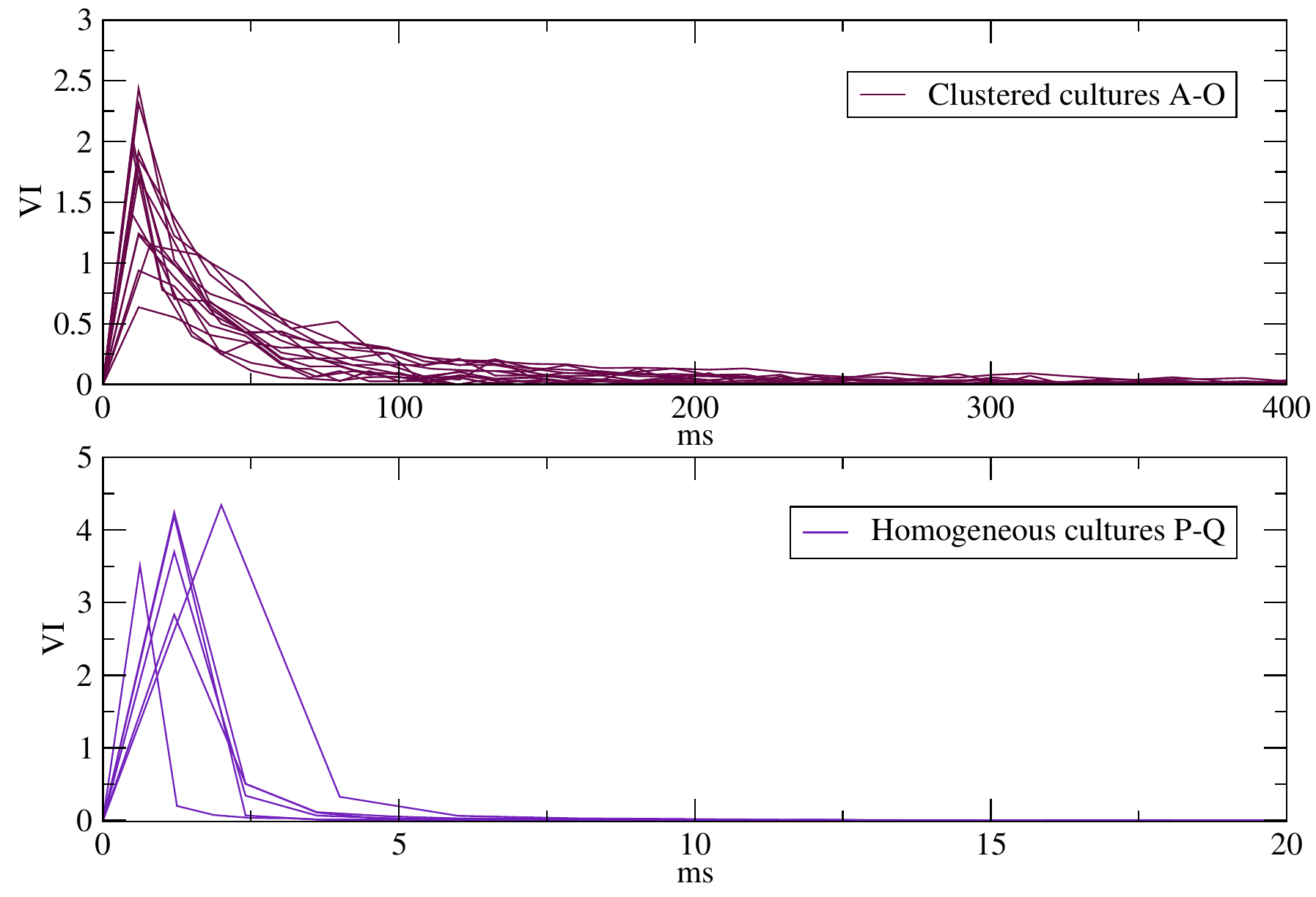}
\end{center}
\caption{
\textbf{Figure S3: Sensitivity of the functional network
construction to the cut--off times.} The cut--off determines
the end of a sequence and therefore its variation modifies the
set of burst chosen. The cut--off is set to $200$ ms for
clustered cultures and $10$ ms for homogeneous ones. To assess
the sensitivity of the grouping of bursts to the cut--off, we
have computed the variation of information between the grouping
of bursts at a certain cut--off value and the previous one. The
variation of information is computed as $VI(X,Y) = H(X) + H(Y)- 2 I(X,Y)$, where $X$ and $Y$ are two partitions, $H$ is the
entropy and $I$ is the mutual information. In our case, each
partition is the set of burst found at a certain value of the
cut--off. We have screened the cut-off values from 0 to 1000
ms. The analysis shows that $10$ ms for homogeneous and $200$
ms for clustered cultures are values for which the Variation of
Information is already stabilized. Thus, the modification of the cut--off within
the stabilization region does not change the grouping of clusters in each burst, and therefore the derived functional networks, as well as the corresponding network measures, remain the same.
}\end{figure}
\newpage

\begin{figure}[!t]
\begin{center}
\includegraphics[width=12cm]{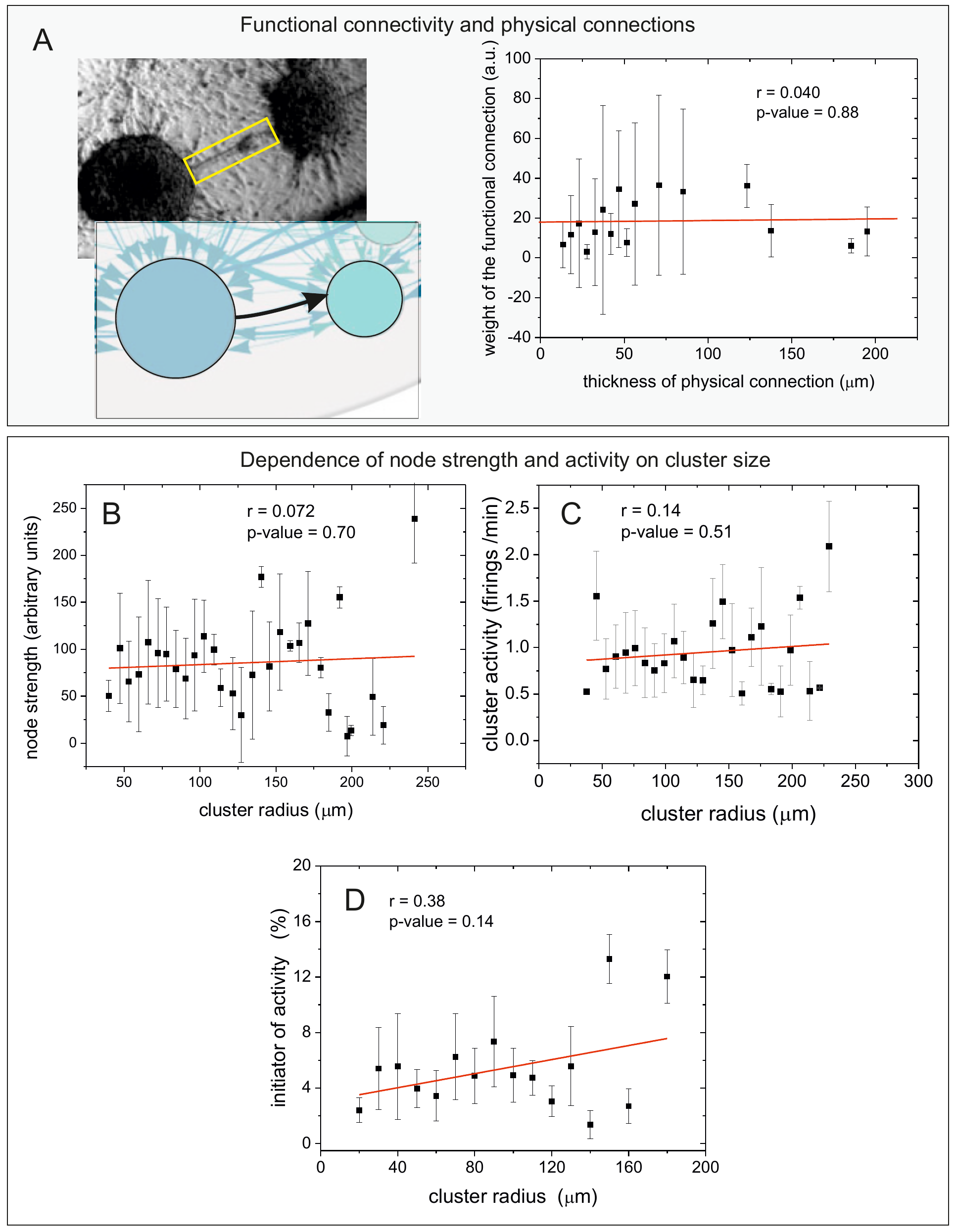}
\end{center}
\caption{
\textbf{Figure S4: Relation between the structural traits of the
network and the functional connectivity.} \textbf{A} Dependence of the
weight of the functional connections on the width of physical
connections between directly connected clusters. The sketch
conceptually shows the comparison between structural and
functional links. The plot represents the analysis of $102$
pairs of clusters, with data binned for similar widths. No
significant correlation is observed. \textbf{B} The dependence of the
node strength on cluster size shows no correlation, indicating
that the functional connectivity cannot be assessed from the
size of the clusters. Data is based on the analysis of $537$
clusters. \textbf{C} For the same clusters, this plot shows that the activity of a cluster
is independent of its size. \textbf{D} Activity within a burst is always initiated by a particular cluster, which triggers the sequential activation of all the downstream clusters. To quantify the importance of these `initiators of activity' in network dynamics
we computed the number of times that a cluster of a given size initiates a
sequence of activations. The plot shows that there is no a significant
correlation between initiation and size. The analysis is based on the study of $1800$ bursts. All these results indicate that the functional connectivity cannot
be drawn from a visual inspection of the neuronal culture.
Errors bars show standard deviation.
}\end{figure}
\newpage

\begin{figure}[!t]
\begin{center}
\includegraphics[width=15cm]{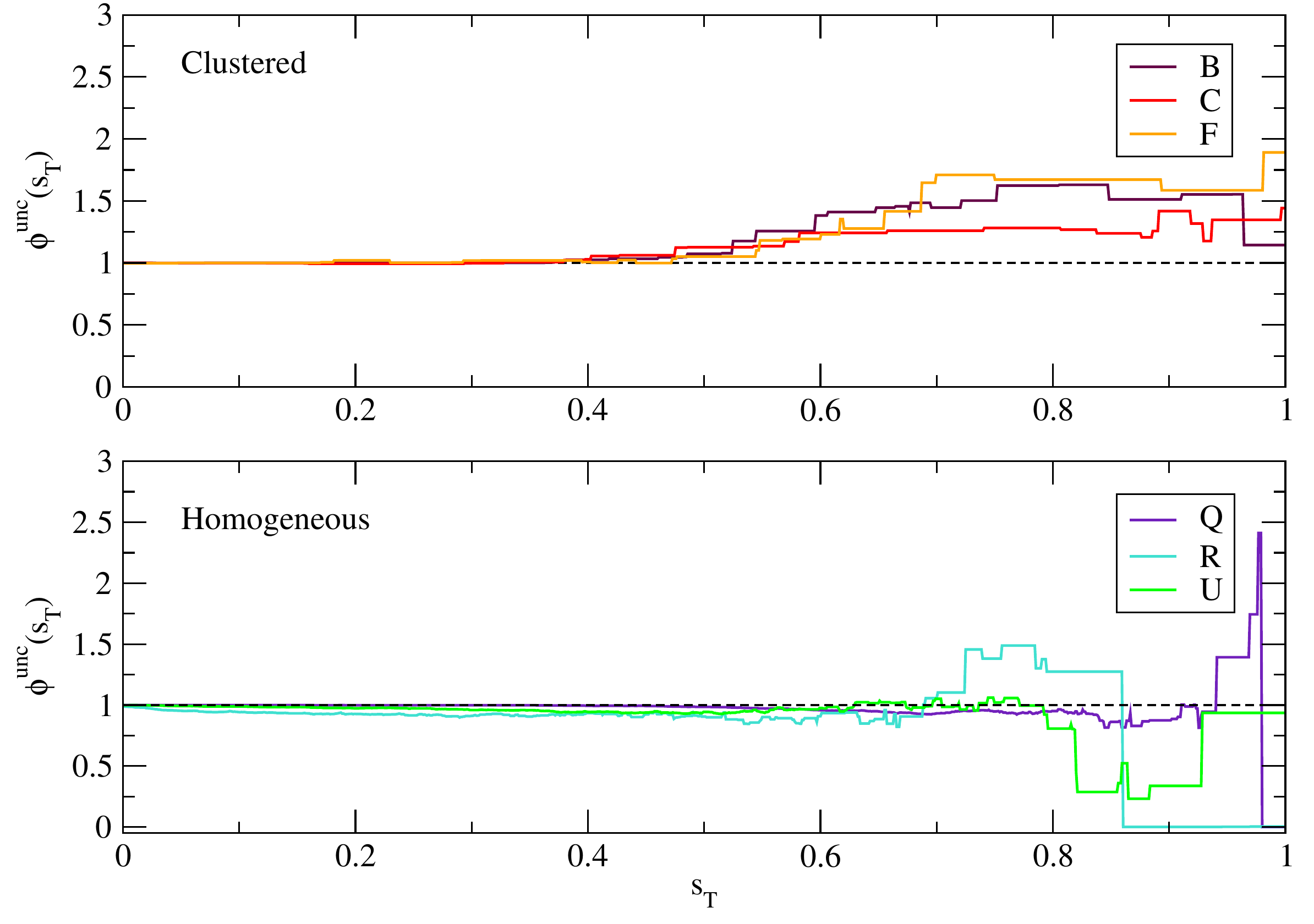}
\end{center}
\caption{
\textbf{Figure S5: `Rich--club' analysis.} The evaluation of the
rich--club $\phi^{\mathrm{unc}}(s_T)$ is performed by computing
the ratio between the  connectivity strength of highly connected
nodes and its randomized counterpart, $\phi^{\mathrm{unc}}=
W_{s_T}/W^{\mathrm{unc}}_{s_T}$, and for gradually larger
values of the strength threshold $s_T$. The figure shows the
rich-club analysis for 3 representative clustered and
homogeneous cultures. Clustered networks exhibit values of
$\phi^{\mathrm{unc}}$ systematically higher than $1$ for large
values of the strength threshold $s_T$, evidencing the
existence of a rich--club core of highly connected clusters in
the network. On the contrary, homogeneous cultures display a
mixture of positive and negative values, and with an average
around $0$, ruling out the existence of the rich-club property.
}\end{figure}
\newpage

\begin{figure}[!t]
\begin{center}
\includegraphics[width=15cm]{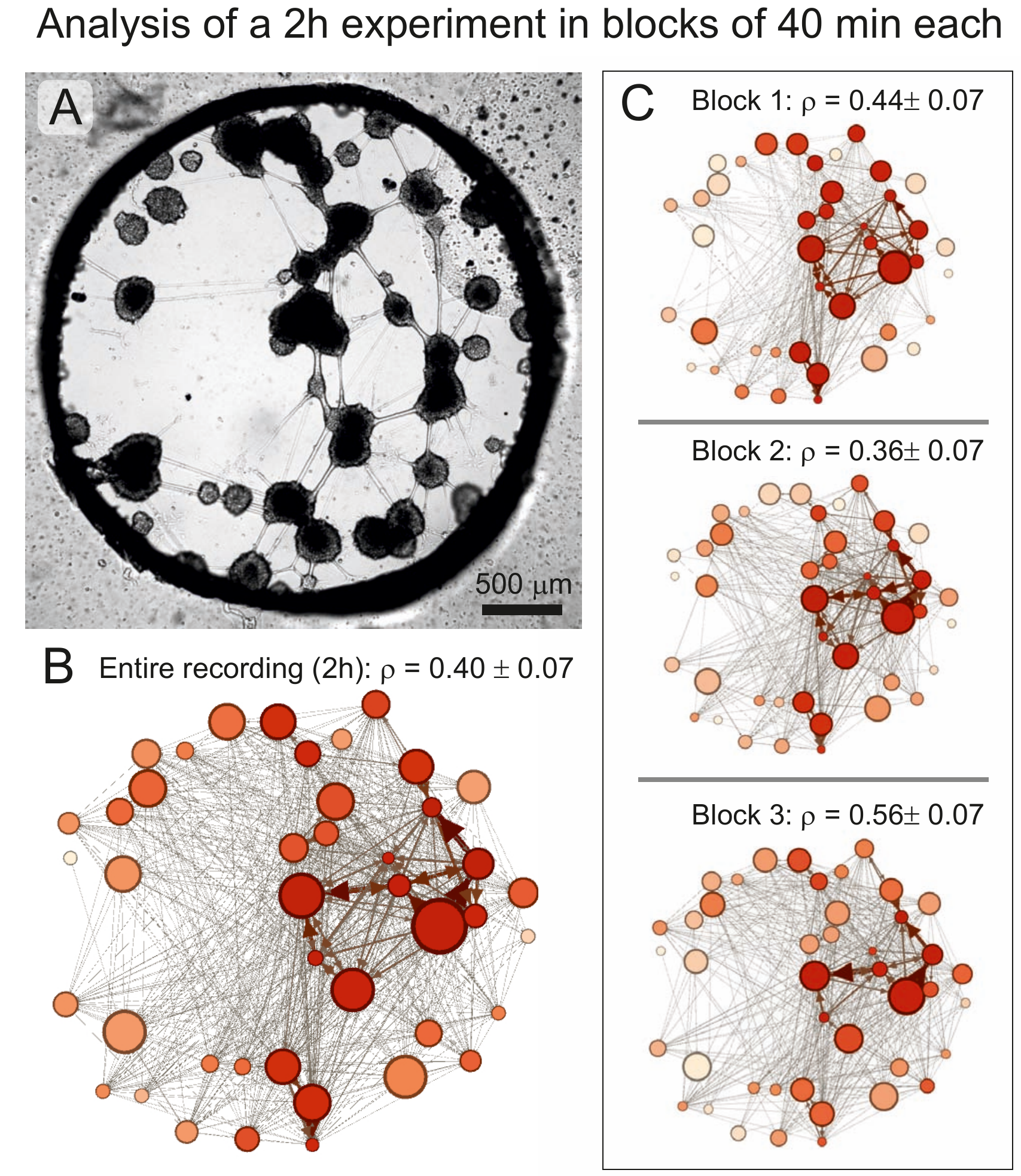}
\end{center}
\caption{
\textbf{Figure S6: Control experiment.} \textbf{A} Bright field image of a clustered network whose spontaneous activity has been recorded for 2h. The average bursting rate of the network is 1.12 bursts/min. \textbf{B} Corresponding functional network. The size of the nodes is proportional to the size of the actual clusters, and their color is proportional to their strength. The weights of the links are both color and thickness coded. The darker the color, the higher the value of the observable. \textbf{C} Analysis of the 2h recording in three blocks, 40 min in duration each, and containing 45 bursts. The blocks show very similar traits between them, as well as with the entire recording. The blocks exhibit similar assortativity values, and share both the most important links and nodes' strengths. $\rho$ indicates the assortativity value of the depicted network, averaged over the Pearson and Spearman formulations.
}\end{figure}
\newpage


\newpage

\end{document}